\documentclass[sn-standardnature]{sn-jnl}
\usepackage{xspace}

\jyear{2021}%

\theoremstyle{thmstyleone}%
\theoremstyle{thmstyletwo}%

\theoremstyle{thmstylethree}%

\newcommand{\kms}{\ensuremath{\mathrm{km\,s^{-1}}}\xspace}

\begin{document}

\title[A small and vigorous black hole in the early Universe]{A small and vigorous black hole in the early Universe}


\author*[1,2,3]{\fnm{Roberto} \sur{Maiolino}}\email{rm665@cam.ac.uk}

\author[1,2]{\fnm{Jan} \sur{Scholtz}}
\author[1,2]{\fnm{Joris} \sur{Witstok}}
\author[4]{\fnm{Stefano} \sur{Carniani}}
\author[1,2]{\fnm{Francesco} \sur{D'Eugenio}}
\author[5]{\fnm{Anna} \sur{de Graaff}}
\author[1,2]{\fnm{Hannah} \sur{\"{U}bler}}
\author[1,2]{\fnm{Sandro} \sur{Tacchella}}
\author[6]{\fnm{Emma} \sur{Curtis-Lake}}
\author[7]{\fnm{Santiago} \sur{Arribas}}
\author[8]{\fnm{Andrew} \sur{Bunker}}
\author[9]{\fnm{Stéphane} \sur{Charlot}}
\author[8]{\fnm{Jacopo} \sur{Chevallard}}
\author[10]{\fnm{Mirko} \sur{Curti}}
\author[1,2]{\fnm{Tobias~J.} \sur{Looser}}
\author[11]{\fnm{Michael~V.} \sur{Maseda}$^\text{11}$}
\author[12]{\fnm{Tim} \sur{Rawle}}
\author[7]{\fnm{Bruno} \sur{Rodríguez Del Pino}}
\author[13]{\fnm{Chris~J.} \sur{Willott}}
\author[14]{\fnm{Eiichi} \sur{Egami}}
\author[15]{\fnm{Daniel} \sur{Eisenstein}}
\author[14]{\fnm{Kevin} \sur{Hainline}}
\author[16]{\fnm{Brant} \sur{Robertson}}
\author[17]{\fnm{Christina~C.} \sur{Williams}}
\author[14]{\fnm{Christopher~N.~A.} \sur{Willmer}}
\author[1,2]{\fnm{William~M.} \sur{Baker}}
\author[18,19]{\fnm{Kristan} \sur{Boyett}}
\author[14]{\fnm{Christa} \sur{DeCoursey}}
\author[20]{\fnm{Andrew~C.} \sur{Fabian}}
\author[14]{\fnm{Jakob~M.} \sur{Helton}}
\author[14]{\fnm{Zhiyuan} \sur{Ji}}
\author[8]{\fnm{Gareth~C.} \sur{Jones}}
\author[21]{\fnm{Nimisha} \sur{Kumari}}
\author[1,2]{\fnm{Nicolas} \sur{Laporte}}
\author[22]{\fnm{Erica} \sur{Nelson}$^\text{22}$}
\author[7]{\fnm{Michele} \sur{Perna}}
\author[1,2]{\fnm{Lester} \sur{Sandles}}
\author[14]{\fnm{Irene} \sur{Shivaei}}
\author[14]{\fnm{Fengwu} \sur{Sun}}

\affil[1]{Kavli Institute for Cosmology, University of Cambridge, Madingley Road, Cambridge, CB3 OHA, UK}

\affil[2]{Cavendish Laboratory - Astrophysics Group, University of Cambridge, 19 JJ Thomson Avenue, Cambridge, CB3 OHE, UK}

\affil[3]{Department of Physics and Astronomy, University College London, Gower Street, London WC1E 6BT, UK}

\affil[4]{Scuola Normale Superiore, Piazza dei Cavalieri 7, I-56126 Pisa, Italy}

\affil[5]{Max-Planck-Institut f\"ur Astronomie, K\"onigstuhl 17, D-69117, Heidelberg, Germany}

\affil[6]{Centre for Astrophysics Research, Department of Physics, Astronomy and Mathematics, University of Hertfordshire, Hatfield AL10 9AB, UK}

\affil[7]{Centro de Astrobiolog\'ia (CAB), CSIC–INTA, Cra. de Ajalvir Km.~4, 28850- Torrej\'on de Ardoz, Madrid, Spain}

\affil[8]{Department of Physics, University of Oxford, Denys Wilkinson Building, Keble Road, Oxford OX1 3RH, UK}

\affil[9]{Sorbonne Universit\'e, CNRS, UMR 7095, Institut d'Astrophysique de Paris, 98 bis bd Arago, 75014 Paris, France}

\affil[10]{European Southern Observatory, Karl-Schwarzschild-Strasse 2, 85748 Garching, Germany}

\affil[11]{Department of Astronomy, University of Wisconsin-Madison, 475 N. Charter St., Madison, WI 53706 USA}

\affil[12]{European Space Agency, Space Telescope Science Institute, Baltimore, Maryland, US}

\affil[13]{NRC Herzberg, 5071 West Saanich Rd, Victoria, BC V9E 2E7, Canada}

\affil[14]{Steward Observatory University of Arizona 933 N. Cherry Avenue Tucson AZ 85721, USA}

\affil[15]{Center for Astrophysics - Harvard \& Smithsonian, 60 Garden St., Cambridge MA 02138 USA}

\affil[16]{Department of Astronomy and Astrophysics, University of California, Santa Cruz, 1156 High Street, Santa Cruz, CA 95064, USA}

\affil[17]{NSF’s National Optical-Infrared Astronomy Research Laboratory, 950 North Cherry Avenue, Tucson, AZ 85719, USA}

\affil[18]{School of Physics, University of Melbourne, Parkville 3010, VIC, Australia}

\affil[19]{ARC Centre of Excellence for All Sky Astrophysics in 3 Dimensions (ASTRO 3D), Melbourne, VIC, Australia}

\affil[20]{Institute of Astronomy, University of Cambridge, Madingley Road, Cambridge CB3 0HA, UK}

\affil[21]{AURA for European Space Agency, Space Telescope Science Institute, 3700 San Martin Drive, Baltimore, MD 21210, USA}

\affil[22]{Department for Astrophysical and Planetary Science, University of Colorado, Boulder, CO 80309, USA}


\abstract{
Multiple theories have been proposed to describe the formation of black hole
seeds in the early Universe and to explain the emergence of very massive black
holes observed in the first billion years after Big Bang \citep{
inayoshi+2020,
fan_quasars_2022,
volonteri+2023}. Models consider different seeding and accretion scenarios
\cite{trinca_low-end_2022,Banik19,Singh23,Bennett23}, which require the detection and characterisation of black holes in the first few hundred million years after Big Bang to be validated.
Here we present an extensive
analysis of the JWST-NIRSpec spectrum
of GN-z11, an exceptionally luminous galaxy
at z=10.6, revealing the detection of the [NeIV]$\lambda$2423 and  CII*$\lambda$1335 transitions (typical of Active Galactic Nuclei, AGN), as well as semi-forbidden nebular lines tracing gas densities higher than $\rm 10^{9}~cm^{-3}$, typical of the Broad Line Region of
AGN.
These spectral features indicate that GN-z11 hosts an accreting black hole.
The spectrum also reveals a deep and blueshifted CIV$\lambda$1549 absorption
trough, tracing an outflow with velocity $ 800-1000$~\kms, 
likely driven by the AGN. 
Assuming local virial relations, we derive a black hole
mass of $\rm \log{(M_{BH}/M_{\odot})}=6.2\pm 0.3$, accreting at about 5 times the  Eddington rate.
These properties are consistent with both heavy seeds scenarios, or scenarios envisaging intermediate/light seeds experiencing episodic super-Eddington phases.
Our finding naturally explains the high luminosity of GN-z11 and can also provide an explanation for its exceptionally high nitrogen abundance.
}


\maketitle

GN-z11 was recently observed with JWST. The analysis of the NIRCam images revealed an unresolved nuclear component and a disk-like component with a few 100 pc radius \citep{tacchella_jades_2023}.
A first NIRSpec spectrum was presented in \cite{bunker_jades_2023} who found it to be consistent with star formation,
although the presence of
an AGN was not excluded.
Here we explore the latter scenario by using a deeper spectrum of GN-z11.

\begin{figure}[H]%
\centering
\includegraphics[width=0.8\textwidth]{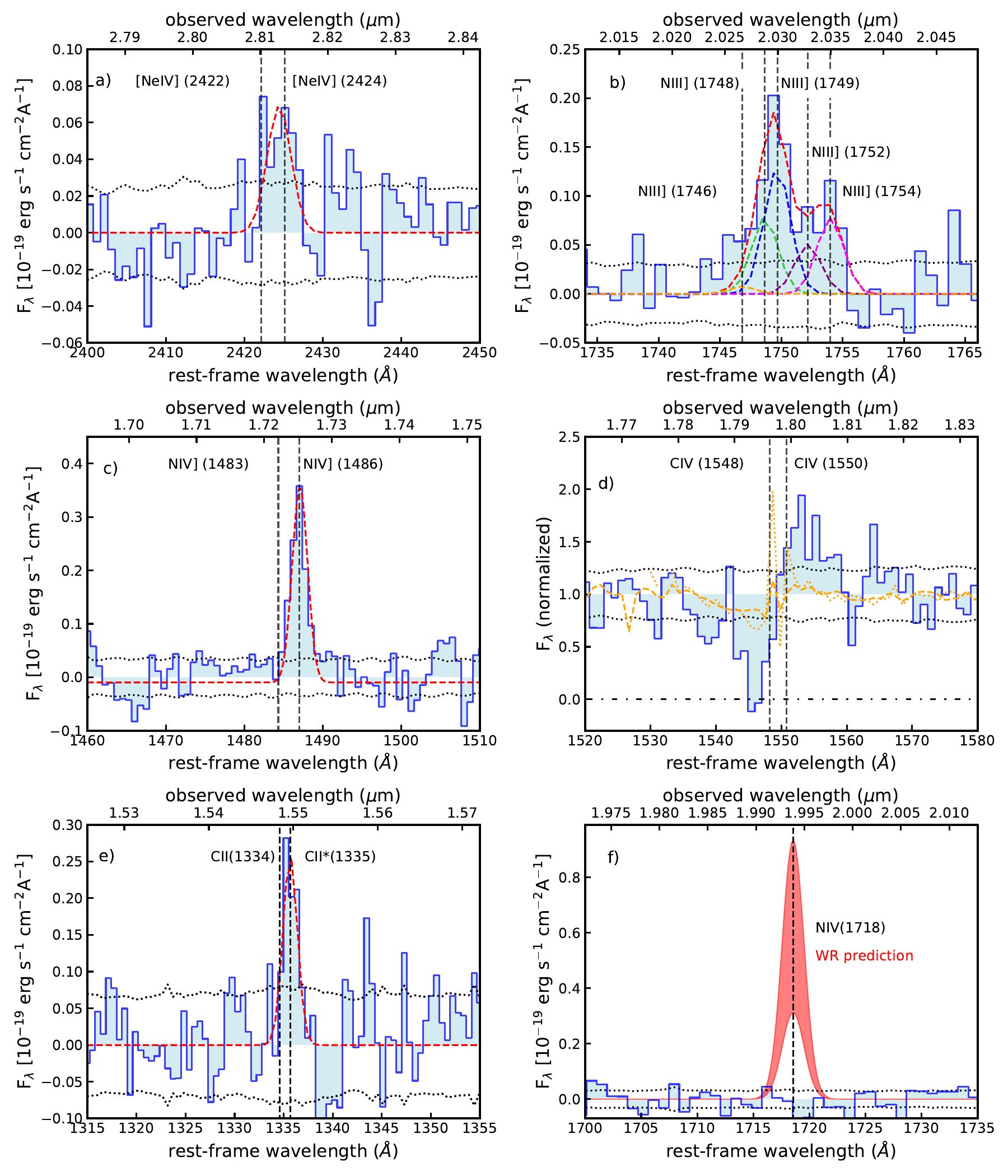}
\caption{Zoom in of the spectra of GN-z11 around specific spectral features of interest, along with their single/multiple Gaussian models (see Methods). Dashed lines indicate the rest-frame wavelengths of the lines at z=10.603.
a) [NeIV]$\lambda\lambda$2422,2424 doublet;
b) NIII] multiplet, illustrating the detection of the resolved NIII]$\lambda$1754 emission;
c) NIV] doublet, showing the absence of [NIV]$\lambda$1483 despite the strong NIV]$\lambda$1486;
d) CIV blueshifted absorption trough and redshifted resonant emission, compared with the CIV P-Cygni profile observed in low-metallicity, young star-forming galaxies (stack: orange dashed line; most extreme case: orange dotted line), showing inconsistency with the latter.
{\b e) CII/CII$^*\lambda\lambda$1334,1335 doublet (seen in emission, without P-Cygni, only in type 1 AGN); f) expected flux of the NIV1718 line in the case that NIV]1486 was associated with WR stars.}
In panels a, b, c, e,  and f the continuum is subtracted, while in panel d the continuum is normalised to one. The grey dotted lines indicate the noise level (1 $\sigma$).}\label{fig:spectra}
\end{figure}

Fig.\ref{fig:spectra}a shows the detection of the [NeIV]$\lambda\lambda$2422,2424 doublet. As NeIV requires photons more energetic than 63.5~eV, this line is an unambiguous AGN tracer \citep{Feltre16,Terao22,Lefevre19} and not seen in star forming galaxies, not even those hosting Wolf-Rayet stars \citep{Hainich14}.

We also detect  CII$^*\lambda$1335 emission (Fig.\ref{fig:spectra}e). This line is commonly observed in AGN \citep{Vanden_Berk01,Wu22,grazian+2020}. In star forming galaxies this line is generally totally undetected; when detected, is extremely faint and always associated with deep CII$\lambda$1334 resonant absorption  \citep{berg+2022},
not seen in GN-z11.

\begin{figure}[h!]%
\centering
\includegraphics[width=0.75\textwidth]{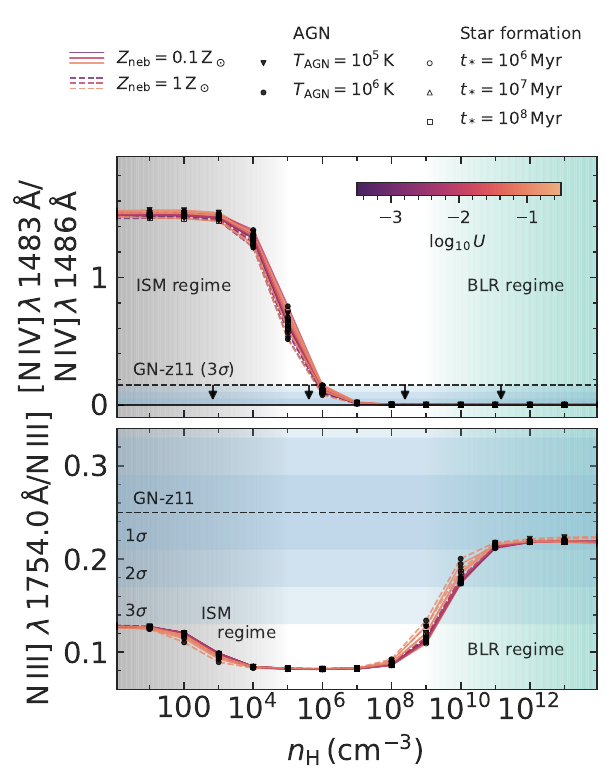}
\caption{
Flux ratios of density-sensitive nitrogen lines as a function of hydrogen gas density, $n_\text{H}$. A large range of Cloudy models (see Methods) are compared with the values observed in GN-z11. Models with metal-poor ($Z_\text{neb} = 0.1 \, \mathrm{Z_\odot}$) and metal-rich ($Z_\text{neb} = 1 \, \mathrm{Z_\odot}$) gas are shown with solid lines and dashed lines, respectively, (color-coded according to the ionization parameter $U$) in the scenario where either an AGN (filled symbols demarcating different black body temperatures for the accretion disc, $T_\text{AGN}$) or stellar populations (open markers for various ages, $t_*$) is responsible for the incident radiation field. Top: [NIV]$\lambda$1483/NIV]$\lambda$1486 flux ratio. Bottom: ratio of NIII]$\lambda$1754 to total flux of the multiplet. The black dashed lines and blue shaded regions (in decreasing darkness for $1\sigma$, $2\sigma$, and $3\sigma$ confidence level as indicated) show the observed fractional contribution of NIII]$\lambda$1754 and upper limit on [NIV]$\lambda$1483/NIV]$\lambda$1486 obtained for GN-z11, indicating that the gas emitting these lines has high density  ($n_\text{H} \gtrsim 10^{9} \, \mathrm{cm^{-3}}$ at $3\sigma$).
The light green shaded areas highlight the range of densities typical of the Broad Line Regions (BLRs), while the gray shaded regions highlight the range of densities typical of the ionized ISM.
}\label{fig:cloudy}
\end{figure}

The [NIV]$\lambda$1483, NIV]$\lambda$1486 doublet is very sensitive to the gas density (while being  insensitive to ionization parameter, metallicity and shape of the ionizing spectrum). Fig.\ref{fig:spectra}c shows the detection of the semi-forbidden NIV]$\lambda$1486 line (critical density $\rm 4.7\times 10^9~cm^{-3}$) and the non-detection of the forbidden [NIV]$\lambda$1483 line (critical density $\rm 1.5\times 10^5~cm^{-3}$), which indicate densities much higher than $\rm 10^5~cm^{-3}$.
Specifically, the various photoionization models shown in Fig.~\ref{fig:cloudy}a (see Methods for details) illustrate that the upper limit on the doublet ratio 
requires densities $\rm \gtrsim 10^6~cm^{-3}$,
which are incompatible with
the densities of the ionized interstellar medium (ISM) that are typically in the range $\rm
10-10^3~cm^{-3}$, and only rarely approach a few times 10$^4$~cm$^{-3}$
\citep{mingozzi_classy_2022}.

Even stronger constraints come
from the NIII] multiplet (Fig.\ref{fig:spectra}b). This is
contributed primarily by four semi-forbidden lines at 1748.6\AA, 1749.7\AA, 1752.2\AA\ and 1754.0\AA \ (and the much weaker 1746.8\AA).
All of these have high critical
densities ($\rm > 10^9~cm^{-3}$), but the 1748.6\AA\ and 1754.0\AA \ transitions have the highest critical density of $\rm 10^{10}~cm^{-3}$
(note that atomic physics requires fixed flux ratios $F_{1754}/F_{1748}=1.05$ 
and $F_{1746}/F_{1752}=0.14$, see Methods). 
The 1754.0\AA\ line is well resolved from the rest of the multiplet, and
its intensity is well constrained to be 0.25$\pm$0.04 of the total intensity of
the multiplet. Such a high ratio can only be achieved when both the 1749.7\AA\ and 1752.2\AA\ transitions are suppressed relative to the  1748.6\AA\ and 1754.0\AA\ because of the very high
density. As illustrated in Fig.\ref{fig:cloudy}b, 
when compared with the expectations from a wide range of photoionization models, the observed ratio requires a density  higher than 10$^{9}$~cm$^{-3}$ at 3$\sigma$ (higher than
10$^{10}$~cm$^{-3}$ at 2$\sigma$). Such high densities are completely inconsistent with any HII regions in any star-forming galaxy, but are fully in the realm of the Broad Line Regions (BLRs) of AGN,
which are indeed characterized by extremely high densities ($\rm \sim
10^{9}-10^{15}~cm^{-3}$).

Therefore, the most plausible explanation is that GN-z11 hosts an AGN and that these semi-forbidden lines observed in its  spectrum are mostly emitted by the associated BLR.

It may appear puzzling that 
the spectrum of GN-z11 does not seem to show the
typical `broad lines' seen in type 1 quasars and AGN, with widths of thousands \kms. However, the width of the
broad lines scales quadratically with the black hole mass, hence in the case of
`small' black holes the broadening is expected to be significantly
smaller.
Moreover, there are classes of type 1 AGN that have broad lines with widths $<$1,000~\kms: these
are the so-called Narrow Line Seyfert 1 (NLSy1), whose permitted lines are
broader than their forbidden lines, but not by a large factor, and in many cases reaching a width of only a few 100 \kms
\citep{osterbrock+pogge1985},
and which are indeed inferred to have small black holes ($\sim 10^6~M_\odot$) 
\citep{Mathur12}.
This seems to be the case of GN-z11, whose semi-forbidden (NIII], NIV]) and permitted lines (MgII) all have widths between 430 \kms and 470 \kms, while [NeIII] has a significantly narrower width (340$\pm$30~\kms), hence coming from the host galaxy (either from HII regions or from the Narrow Line Region of the AGN).
We note that the intepretation of some permitted and semi-forbidden lines, such as the Balmer lines and CIII] is made complex by the fact that these  are also generally contributed to by the ISM photoionized by star formation in the host galaxy.

The spectrum of GN-z11 also reveals a 
deep (EW$_{rest}$$\sim$5\AA) and blueshifted absorption trough of the CIV$\lambda\lambda$1548,1550 doublet (Fig.\ref{fig:spectra}d).
Deep CIV  absorption is
sometimes observed in young stellar populations, but the depth observed in GN-z11 would require
high metallicities, typically solar or super-solar 
\citep{leitherer+2011}. This is in contrast with
the metallicity inferred from the nebular lines of GN-z11
($\rm Z\approx 0.1Z_\odot$) \citep{bunker_jades_2023}.
To illustrate more quantitatively the inconsistency with the stellar-wind origin, the orange dashed line in Figure \ref{fig:spectra}d shows the stacked spectrum of local galaxies with metallicity around the value inferred for GN-z11,
resampled to the NIRSpec grating resolution: the stellar trough is much shallower than observed in GN-z11 and with a completely different shape.
Aside from the stellar origin, such a deep CIV absorption is seen also in lower redshift star forming galaxies and associated with galactic outflows
\citep{Du16}.
However, in those cases the outflow velocities are only of a few 100 \kms (see Methods), while in the case of GN-z11 the CIV trough traces a much faster outflow of $\rm \sim 800-1000~\kms$.
A more plausible explanation of the deep blueshifted trough of CIV is that GN-z11
is a Broad Absorption Line (BAL) quasar, which are indeed characterised by deep
 absorption of blueshifted CIV by up to several thousands \kms. Actually, GN-z11 would fit in the `mini-BAL' category, with velocities between 500 and 2000 \kms, more common in lower luminosity AGN,
 or in the ``Narrow'' ($\sim$1,000~km/s) Absorption Line (NAL) quasars category \citep{Elvis00}.
The spectrum also shows a clear CIV redshifted emission, which is likely tracing the receding component of the outflow. Indeed, since CIV is a resonant line, this is  the counterpart of the redshifted Ly$\alpha$ identified by \cite{bunker_jades_2023}  (consistent shift and width).

Summarizing, the detection of [NeIV] and CII$^*$, the extremely high gas density matching those of the AGN BLRs, and the presence of a deep, blueshifted absorption trough of CIV tracing a high velocity outflow, are all consistent with the scenario in which GN-z11 hosts an accreting black hole, i.e. an AGN, specifically what would be called NLSy1 and (mini-)BAL/NAL AGN.

In the Methods we also discuss other diagnostics, such as the ratio of UV transitions (e.g. CIII]/CIV, CIII]/HeII) and the upper limits on high ionization lines (NV$\lambda\lambda$1238,1242  and [NeV]$\lambda$3426), are fully consistent with the AGN scenario.

Some works have suggested that GN-z11 may host a population of WR stars
\citep{Senchya23}. The HeII$\lambda$1640 line shows a potentially broad profile, as illustrated in Fig.\ref{fig:spectra_appendix}c of the Methods (although the wings are mostly in the noise). This, if confirmed, could come from the inner region of the BLR but could also trace the presence of a WR population. However, various other features are inconsistent with a major contribution from WR stars. Specifically, in the case of WR stars the NIV$\lambda$$\lambda$1483,1486 doublet, if present, is always accompanied by an even stronger NIV$\lambda$1718 line, with a prominent P-Cygni profile, which is not seen at high confidence in the spectrum of GN-z11 \citep{Hainich14} (Fig.\ref{fig:spectra}f); [NeIV] and CII$^*$ are never seen associated with WR stars \citep{Hainich14}; when present, the NIII] multiplet has a much weaker $\lambda$1754 component \citep{mingozzi_classy_2022}. Therefore, if WR stars are present in GN-z11, then they must co-exist with the AGN and are unlikely to play a dominant role in the excitation of the observed nebular lines.

\begin{figure}[h!]%
\centering
\includegraphics[width=0.9\textwidth]{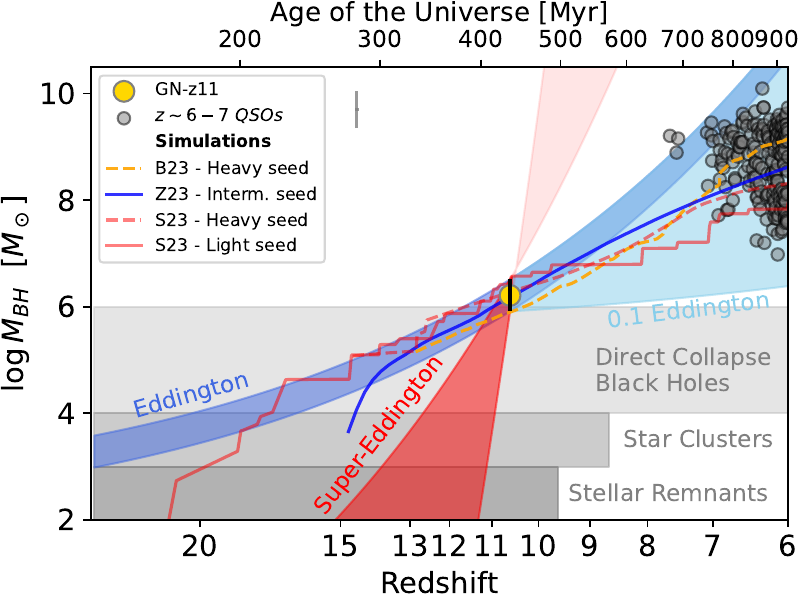}
\caption{Black hole mass as a function of redshift (on a logarithmic scale) and age of the Universe. The black hole mass inferred for GN-z11 is shown with the large golden symbol.
The red shaded region indicates the evolution expected in the case of super-Eddington accretion at the level inferred for GN-z11. The darker blue shaded region shows the black hole mass evolution assuming Eddington-limited accretion, while the lighter blue shaded region shows the case of evolution in the case of sub-Eddington accretion (between 0.1 and 1 the Eddington rate).  The horizontal gray shaded regions indicate the range of black hole seeds expected by different scenarios. Solid and dashed lines indicate the evolutionary tracks of various simulations and models \cite{Bennett23,Zhang_Trinity_23_IV,
Schneider23} that can reproduce the GN-z11 BH mass, with different seeding and accretion rate assumptions, as detailed in the Methods. The small gray symbols indicate the black holes measured in quasars at z$\sim$6--7.5 \citep{inayoshi+2020,fan_quasars_2022}  (whose representative 1$\sigma$ errorbar is shown in the top-left), most of which can originate from a progenitor like the black hole in GN-z11.} \label{fig:mbh_z}
\end{figure}

Assuming local virial relations, 
the black hole mass can be estimated from the line widths and continuum luminosity. As discussed in the Methods, we estimate a BH mass of about $\rm 1.6  \times 10^6~M_{\odot}$. 
 In the Methods we also discuss potential uncertainties and caveats in the determination of the BH.

We infer a bolometric luminosity of the AGN of $\rm 10^{45}~\rm erg/s$ (see Methods), which is a factor of $\sim$5 higher than the Eddington limit (with an uncertainty of a factor of 2). Super-Eddington accretion is generally inferred for NLSy1's and is one of the scenarios proposed to rapidly grow supermassive black holes in the early Universe 
\citep{trinca_low-end_2022,Schneider23}.

 Fig.\ref{fig:mbh_z} shows how the black hole mass in GN-z11 would have evolved at earlier cosmic epochs if accreting at the Eddington rate, or at the super-Eddington rate estimated at the time of observation. For comparison, the gray shaded areas show the range of possible black hole seeds scenarios:
black holes resulting
from the direct collapse of primordial clouds
into seeds with masses in the range $\sim 10^4-10^6~M_\odot$, the
so-called Direct Collapse Black Holes (DCBHs);
rapid merging of stars and black holes in dense, nuclear star clusters; accretion onto Population III black hole
seeds or even normal stellar remnants \citep{inayoshi+2020,Volonteri23,trinca_low-end_2022,Banik19,Singh23}. Many of these semianalytical models and cosmological simulations could reproduce the mass of GN-z11 at z=10.6 \cite{
zhu+2020,
bhowmick+2022,
Bennett23,
Schneider23}. The solid and dashed lines show the evolutionary tracks for some of them (described more extensively in the Methods). These can be broadly divided in models assuming heavy seeds (DCBH), whose accretion is limited to the Eddington rate, and intermediate mass (stellar clusters) or light (stellar remnants) seeds experiencing episodes of super-Eddington accretion. It is interesting to also note that GN-z11 evolving at sub-Eddington rate can easily result into the super-massive black holes ($\rm 10^7-10^9~M_\odot$) observed in quasars at z=6-7, as indeed predicted by many models.

However, it is possible that the local/low-z scaling relations do not apply for AGN at such early epochs. If we disregard the local virial relations and instead assume that the black hole in GN-z11 is accreting at the Eddington rate, then the black hole mass would be $10^7~M_\odot$. A black hole with this mass is more difficult to account for, but achievable in models assuming heavy seeds and episodes of super-Eddington accretion \cite{Bennett23,zhu+2020,
bhowmick+2022}.

\begin{figure}[h!]%
\centering
\includegraphics[width=0.9\textwidth]{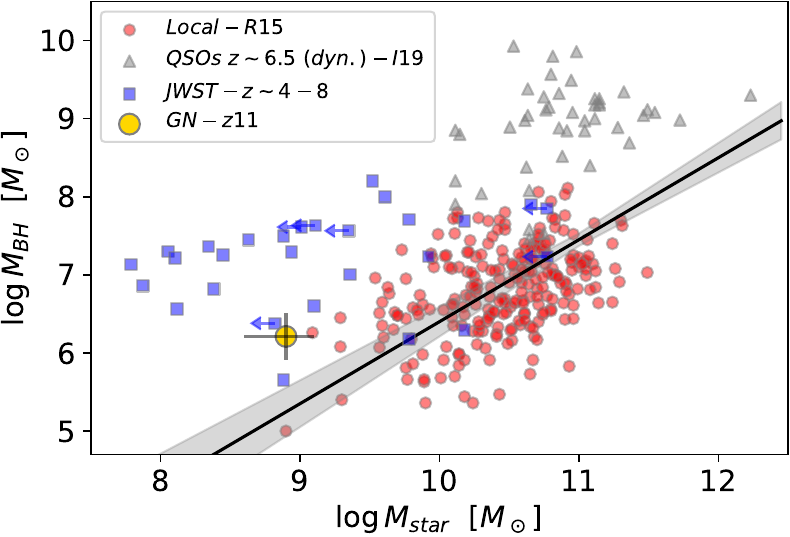}
\caption{
Black hole versus stellar mass diagram, showing the
location of GN-z11 (large golden symbol), compared to local galaxies as indicated by the small red symbols and their best-fit relation (black solid line and uncertainty traced by the gray shaded region) \citep{Reines15}. The grey symbols show the values estimated for  quasars at z$\sim 6-7$ \citep{izumi_subaru_2019}, although in these cases the galaxy mass is inferred from dynamical tracers. The blue symbols are AGN at z$>$4 for which the black hole and galaxy stellar mass has been measured with JWST data  (see Methods) using the same calibration as \citep{Reines15} for consistency.
}\label{fig:mbh_mst}
\end{figure}

Taking the stellar mass of the extended disk-like component measured by \cite{tacchella_jades_2023} ($M_{\star}=8\times10^8~M_\odot$), 
it is possible to locate GN-z11 on the $M_{BH}-M_{star}$ relation. As illustrated in Fig.\ref{fig:mbh_mst},
GN-z11 is placed above the local relation, although marginally consistent within the scatter. Interestingly, an early evolution above the local $M_{BH}-M_{star}$ relation is what is expected from models invoking DCBHs and/or super-Eddington accretion \citep{trinca_low-end_2022, Volonteri23, Schneider23}.

We note that the exceptionally high nitrogen abundance inferred for GN-z11 (specifically, high N/O) \citep{bunker_jades_2023}
becomes much less problematic in the AGN scenario. 
To begin with, several `nitrogen-loud' AGN have already been found both at low and high-redshifts
(including NLSy1) 
\citep{
matsuoka+2017, Koptelova2022, ubler+2023, Jiang08}.
So GN-z11 is not a very peculiar system in this context. Secondly, the mass of the BLR in AGN is very small \cite{Osterbrock06}:
$$
\rm M_{BLR} = 1.8 \left( \frac{L_{H\gamma}}{10^{42}~\rm erg/s}\right)
\left( \frac{n}{10^{10}~\rm cm^{-3}}
\right)^{-1} ~\rm M_\odot
$$
Specifically, the H$\gamma$ luminosity observed in GN-z11 ($L_{H\gamma}= 1.7\times 10^{42}~erg/s$, even assuming the extreme case it is not contributed by the NLR and HII regions) implies a mass of the BLR of only a few solar masses.
It would take just one or two supernovae to enrich such a small mass to solar/super-solar metallicity, 
especially  within the small physical region associated with the BLR ($\sim$ 10$^{-2}$~pc for GN-z11)
\cite{Mandal21}.
These could be supernovae from supermassive stellar progenitors, with high nitrogen yields \citep{volpato+2023}.
However, even without invoking exotic scenarios, given the accelerated metal enrichment of such a small, central region, it is also possible that secondary, recycled nitrogen production can occur within a timescale of a few tens Myr (especially given the very fast cooling times at such high densities, which allow star formation to quickly occur out of cooled SN ejecta).

We finally discuss our results within the context of the recent JWST findings of an excess of exceptionally luminous galaxies at high redshift.
GN-z11 is the first of such hyperluminous galaxies at high-z to be spectroscopically confirmed, and for which such a detailed spectroscopic analysis has been feasible. The AGN scenario revealed by our analysis provides a natural explanation for the exceptional lumionsity of GN-z11. If this is representative of the broader class of luminous galaxies discovered at high-z, then it would greatly alleviate the tension with models and simulations.

\vspace{10truecm}
\vskip10truecm

\newpage

\section{Methods}

\subsection{Observations and Data Processing}
The data presented in this paper are part of the JADES survey \citep{Eisenstein23} and, specifically, obtained through programme ID 1181 (P.I.: D. Eisenstein). GN-z11 was observed in two epochs, the first one on UT 5 and 7 February 2023,
and the second one on UT 4 and 5 May 2023.
The February observations were already presented in \cite{bunker_jades_2023}. We refer to that paper for a detailed description.
Briefly, the spectroscopic data were obtained with four different configurations of the NIRSpec micro-shutter array (MSA) \citep{Jakobsen22,Ferruit22,Boker23}, using a 3-shutter nodding pattern. Four different dispersers were used to cover the 0.6-5.3 micron wavelength range: the low-resolution prism mode (exposure time of 6200 sec), and three medium-resolution gratings (3100 sec each), which provide a nominal spectral resolution of R$\sim$1000 for a uniformly-illuminated slit \citep{Jakobsen22}. However, the highly compact light profile of GN-z11, with respect to the width of the slit, results in a significantly higher effective resolution. To estimate the effective resolution, we forward model the morphology of GN-z11 through the NIRSpec instrument for the grating dispersers, finding that the resolution ranges between 1100 and 2100. Four MSA configurations were used (two pointing and two dither positions). The May observations were similar, but in this case they consisted in three consecutive dithers (with three different MSA configurations), each with three nods, resulting into an on-source exposure of 2.7 hours for the prism, each of the three medium resolution gratings and also with the high resolution grating G395H/290LP. Unfortunately, at the location of GN-z11 on the MSA the latter spectrum is heavily truncated at wavelengths longer than 4.1$\mu$m, hence all strong optical emission lines are not observed with this grating.

By combining the two sets of observations, the total exposure time is 9.6 hours with the prism and 6.15 hours with each of the medium resolution gratings.

The data processing is also described in \cite{bunker_jades_2023}, and we refer to that paper for a detailed discussion. Here we only mention that we used the pipeline developed by the ESA NIRSpec Science Operations Team and the NIRSpec GTO Team. 
 Most of the processing steps in the pipeline adopt the same algorithms used in the JWST Science Calibration Pipeline \cite{bushouse_2023}. Differently from the official pipeline, the final  1D combined spectra are obtained by combining the 1D individual spectra rather than performing the extraction process in the combined 2D spectra. This step guarantees that the final 1D spectra are well flux calibrated for slit-losses. In the combination process, we also applied a 3$\sigma$-clipping algorithm and excluded bad pixels based on the data quality files provided by the pipeline. The extraction of 1D spectra in the individual exposures is also optimized on the basis of science.
In this paper, we adopt a 3-pixels (0.3$''$) extraction along the slit, as it improves the S/N of the spectrum (for point sources).
 Finally, the GTO pipeline provides spectra beyond the nominal wavelength range for the spectral configuration G140M/F070LP by taking into account the transmission filter throughput in the flux calibration processing step. The extended spectra cover the wavelength range of 1.27 $\mu$m – 1.84 $\mu$m.

Moreover, here we combine the grating spectra in their overlapping ranges, which increases the S/N in those regions. The combined spectra in these regions were resampled to 8\AA \ around the NIV doublet, not to affect resolution, and to 12\AA \ around the CIV, as in this case higher S/N is required on the continuum to properly trace the CIV absorption.

In the paper we adopt  the flat $\lambda$CDM cosmology from Planck18 with H$_0$ = 67.4 km/s/Mpc and $\Omega _m$ = 0.315 \citep{Planck18}.

\subsection{Emission line fitting}

Emission lines were fitted by using one or multiple Gaussian lines and a simple power law for continuum subtraction. The best-fit parameters for the continuum and Gaussian components were found using the MCMC algorithm to estimate the uncertainties. For the purposes of this paper, each line in the rest-frame UV was fitted independently, except for doublets or multiplets, whose line widths were forced to the same value (but see discussion below for the CIII] doublet) and the relative wavelength separation of the doublet/multiplet was forced to the nominal rest-frame wavelength. The absolute velocity of each line (or group of lines in the case of doublets and multiplets) was not constrained to the exact redshift given in \cite{bunker_jades_2023}, in order to both allow for small wavelength calibration uncertainties associated with the positional uncertainties of the target within the shutter\footnote{The uncertainties in the target acquisition may result in the target being offset by up to $\sim 0.05''$ relative to the nominal position, which is used by the pipeline for the wavelength solution. Such unknown offset would introduce a wavelength offset by up to 0.5 spectral pixel, which corresponds to a different velocity offsets (given the wavelength-dependent resolution) in different regions of the spectrum.} (Jakobsen, priv. comm.) and also to allow for small velocity shifts between different lines, which are common in AGN, and especially in the BLR. We restricted our fitting to the lines of interest for this paper. 

\begin{figure}[h]%
\centering
\includegraphics[width=0.9\textwidth]{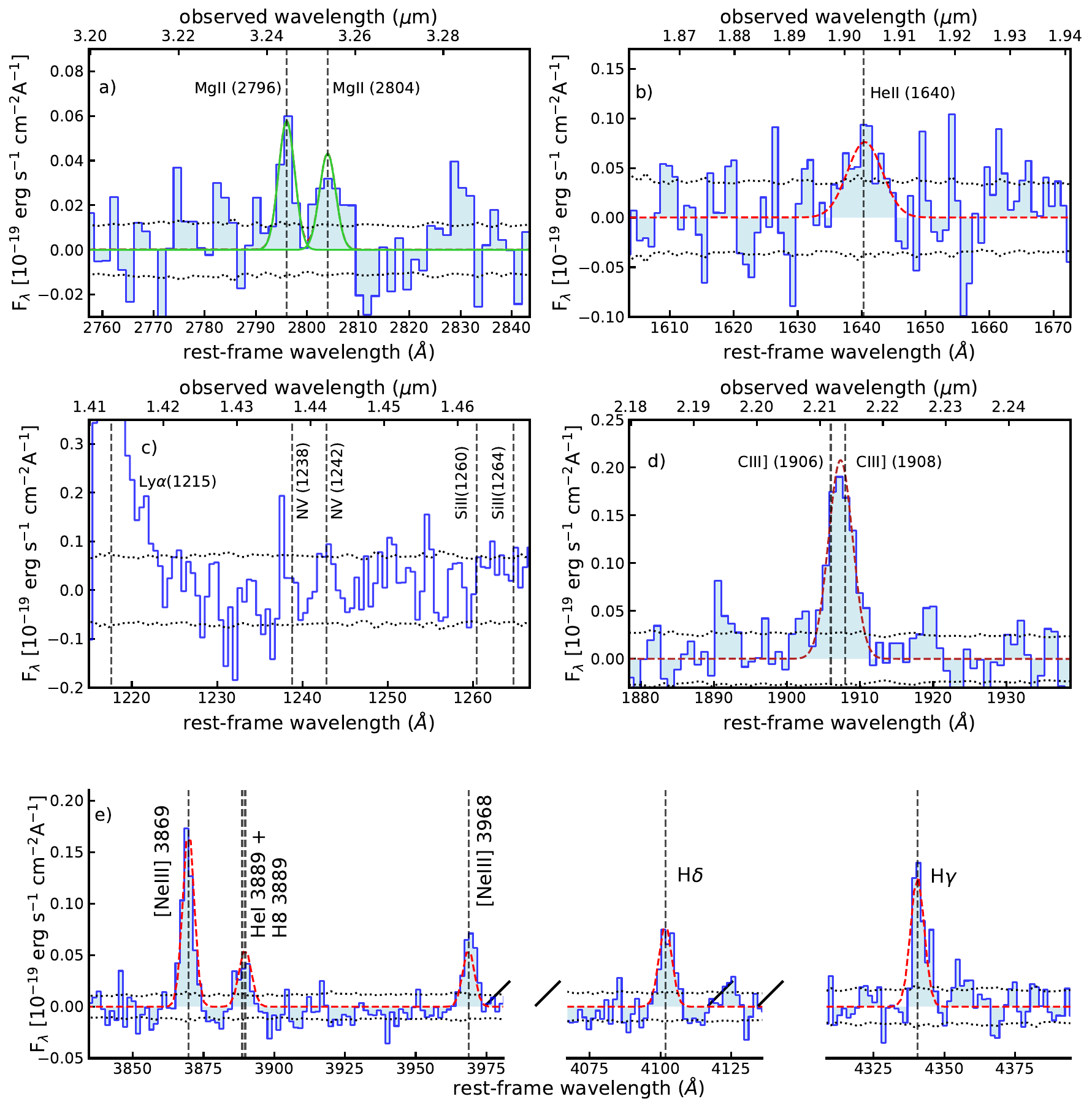}
\caption{Zoom in on the additional emission lines fitted. a) CIII]$\lambda$1906,1908 doublet. As the doublet is unresolved, the fit turns out  degenerate between line width and fluxes of the two components; moreover it is also  contributed to by star formation in the host galaxy (see text for details); b) MgII$\lambda$2796,2804 doublet; c) HeII$\lambda$1640; d) Ly$\alpha$, NV$\lambda$1238,1242 doublet (undetected) and SiII$\lambda$1260,1264 (undetected), corrected for the Ly$\alpha$ damping wing;  e) [NeIII]$\lambda$3869 profile compared with the Balmer lines H$\delta$ and H$\gamma$. In all panels, the continuum is subtracted. The black dotted lines indicate the noise level.}\label{fig:spectra_appendix}
\end{figure}

Table.1 provides the list of the fitted emission line widths and fluxes and Fig. \ref{fig:spectra_appendix} shows the additional fitted lines not shown in the main text.

In the case of the NIII] multiplet, the 
1748.6\AA\ and 1754.0\AA \ transitions
come from the same upper level, hence their flux ratio is fixed by the associated Einstein coefficients, specifically $F_{1754}/F_{1748}=1.05$. Similarly,
1746.8\AA \ and 1752.2\AA\ come from the same upper level and their flux ratio is fixed to $F_{1746}/F_{1752}=0.14$.
The inferred line widths are deconvolved from the line spread function as inferred for the GN-z11 light profile.

For the CIII]1906,1908 doublet it is not possible to resolve the two components; attempting to fit it with two components makes the fit  degenerate between width and intensity of the two components.
The additional caveat of this CIII] doublet is that 
it is also commonly seen in normal star-forming galaxies and in the NLR of AGN, so it can also have a contribution from the host galaxy, as for the [NeIII] emission.  Indeed, \citep{Maiolino23b} use IFS spectroscopy to reveal that the CIII] emission is resolved on scales of several 100 pc. As a consequence we do not include the (spectrally) unresolved CIII] in our analysis, as it does not provide constraints on either the BLR nor the host galaxy.
 In Tab.\ref{tab:emlines} we report the total flux and width using a single Gaussian.
However, for sake of completeness, 
we report that by fitting two components with separate FWHM, accounting for the NLR and BLR, gives CIII]$\lambda$1906/$\lambda$1908 of 0.62$_{-0.37}^{+1.00}$ for the narrow components, with FWHMs of 314$\pm$120 kms$^{-1}$ (consistent with the [NeIII] width), and 560$\pm$80 kms$^{-1}$ for the CIII]$\lambda$1908 broad component (consistent with the NIV] width).

The [OII]3726,3729 doublet would potentially be an additional forbidden line, detected in the observed wavelength range, which could be used to constrain the velocity dispersion in the host galaxy. However, unfortunately, the doublet is unresolved. Attempting to fit it (by forcing the two components to have the same width) results in a FWHM of 365$\pm55$ kms$^{-1}$ and a flux ratio of 0.62$_{-0.21}^{+0.31}$.

\begin{table}
    \centering
    \caption{Measured fluxes and FWHM of the emission lines used in this work.}
    \begin{tabular}{@{}lcccc@{}}
    \hline
    Emission line & $\lambda_{0}$ & Flux & Doublet-Multiplet & FWHM   \\
                  & \AA         &  ($\times 10^{-19}$) ergs s$^{-1}$ cm$^{-2}$ & ratios & km s$^{-1}$ \\
    \hline
    NV       & 1239+1243 & $<$3.5$^a$ & & \\
    $[$NIV$]$    & 1483.32 & $<$ 1.6$^a$ & & \\
    CII* & 1335 & $5.6 _{-1.26} ^{+1.27}$ & & $380\pm80$ \\
    NIV$]$     & 1486.50 & $10.4 _{-1.6} ^{+1.8}$ & &$470\pm50$ \\
    CIV$]$(emission)    & 1550.77&  $3.2 _{-0.8} ^{+0.9}$ & & \\
    HeII    & 1640.50& $6.0 _{-1.3} ^{+1.4}$& & $1200\pm380$ \\
    NIII]   & multiplet-total &
            $10.3^{+2.1}_{-1.3}$ & & $430\pm80^b$ \\
            & 1746.82$^c$/total &  & $0.02^{+0.01}_{-0.01}$ & \\
            & 1748.64$^d$/total &  & $0.24^{+0.05}_{-0.05}$ & \\
            & 1749.67/total &      & $0.37^{+0.06}_{-0.06}$ & \\
            & 1752.16$^c$/total &      & $0.14^{+0.06}_{-0.06}$ & \\
            & 1753.99$^d$/total &  & $0.25^{+0.04}_{-0.04}$ & \\
    CIII]   & 1906+1908 & $9.9 _{-0.9} ^{+0.9}$& & $610\pm60^e$ \\
    MgII    & 2796+2804 & $4.6 _{-0.6} ^{+0.6}$ & & $430\pm65^f$\\   
            & 2796/2804 & &$1.36 _{-0.42} ^{+0.67}$ & \\
    $[$NeV$]$ & 3426.86 & $<1.5^a$  & &  \\
    $[$NeIV$]$ & 2424+2422 (fit)$^g$ & $2.9 _{-0.76} ^{+0.85}$  & &$380\pm110$  \\
           & 2424+2422 (int)$^g$ & $3.14 _{-0.65} ^{+0.65}$  &   \\
    $[$OII$]$& 3727+3729 & $7.72_{-1.30}^{+1.40}$ & & $368\pm50$\\
             & 3729/3727 &  & $0.59_{-0.21}^{+0.30}$& \\
    $[$NeIII$]$ & 3869.68 & $10 _{-0.7} ^{+0.7}$  & &  $340\pm30$ \\
    HeI+H8     & 3889.73  & $3.7 _{-0.75} ^{+0.76}$ & & $360\pm50$ \\
    H$\delta$ & 4101.73 & $5.1 _{-0.75} ^{+0.9}$ & & $410\pm70$ \\
    H$\gamma$ & 4340.47 & $9.2 _{-2.8} ^{+2.9}$& & $360\pm80$ \\ 
    \hline
    \end{tabular}
    \label{tab:emlines}
Notes: $^a$3$\sigma$ upper limit (in the case of NV is the upper limit on the sum of the doublet); $^b$the four components of the multiplet are forced to have the same width in the fit; $^c$as the components $\lambda$1746.82 and $\lambda$1752.16 come from the same level, their fluxes are forced to have the ratio given by their Einstein's coefficients, i.e. 1746.82/1752.16=0.146;
$^d$as the components $\lambda$1748.64 and $\lambda$1753.99 come from the same level, their fluxes are forced to have the ratio given by their Einstein's coefficients, i.e. 1753.99/1748.64=1.05; $^e$for the CIII] doublet is totally blended and is here fitted with a single Gaussian (see text for a more complex, but uncertain decomposition); $^f$the two components of the MgII doublet are forced to have the same width;
$^g$ in the case of the [NeIV] doublet, as it is a key line, we provide both the result from MCMC fitting, assuming a single Gaussian (first line) and also by simply taking the integrated flux (second line).
\end{table}

\subsection{CIV absorption and emission}

In this section, we provide some additional details on the CIV absorption. As mentioned in the text, CIV P-Cygni profiles with a significant CIV blueshifted trough are seen associated with atmospheres of young, hot stars. Yet, the depth of this feature is a strong function of metallicity \citep{leitherer+2011}, and the deep trough observed in GN-z11 would require stars with solar or even super-solar metallicities, completely inconsistent with the much lower metallicity inferred for GN-z11. To illustrate the inconsistency with the pure stellar origin, we have stacked 11  UV spectra from the CLASSY HST survey \citep{berg+2022}, with metallicity around the value inferred for GN-z11 by \cite{bunker_jades_2023} ($\rm Z=0.1~Z_\odot$), specifically 7.4$<$12+log(O/H)$<$7.9. We were conservative by excluding galaxies with strong CIV emission. We also excluded one WR galaxy, as we discuss that the spectrum cannot be dominated by WR stars (see text and section 1.5).   The continuum of the spectra were normalized to one by using a simple linear fit in the spectral ranges 1410--1480\AA\  and 1560--1600\AA , consistent with the analysis of the spectrum of GN-z11 in the same spectral region (Fig.\ref{fig:spectra}d). 
The resulting stacked spectrum is shown with a dashed, orange line in Fig.\ref{fig:spectra}d, and illustrates inconsistency with the trough seen in GN-z11. To be conservative, in Fig.1d we also show the case of the most extreme spectrum among the 11 selected, the one with the deepest CIV absorption. While the wings of the stellar winds can extend out to 2000 km/s, the profile and depth observed trough at these metallicities is inconsistent with the observed trough in GN-z11 at 5$\sigma$.

The blueshifted CIV trough (and redshifted emission) therefore is not a P-Cygni feature associated with stellar (atmosphere) winds. Rather, it is tracing a galactic outflow, as observed in lower redshift starbursts (e.g. \cite{Du16}) and in (mini-)BAL/NAL AGN \citep{Elvis00,Laha21,rodriguez-hidalgo2009,Hamann04,Chen17,Doyee22}. The determination of the velocity requires knowledge of the exact wavelength of the redshifted, rest-frame CIV transition. Unfortunately there are small wavelength uncertainties associated with each grating due to the uncertainties of the location of the sources within the shutter, as discussed above. In this specific case we calibrate the velocity shift based on the NIV line, which is in the same gratings and has similar ionisation potential as CIV. The outflow ``velocity'' is also subject to different definitions. The centroid of the through relative to the mean of the two CIV transitions gives a velocity of --790~km/s. If we consider the blue edge of the trough relative to the bluest of the two transitions (CIV$\lambda$1548.19) then we obtain a velocity of --1040~km/s.  
These velocities are significantly higher than those inferred from the CIV absorption in starburst-driven outflows \citep{Du16,Pettini02}, but in the range of BAL quasars which can span from 500 \kms to several thousands \kms \citep{Elvis00,Laha21,rodriguez-hidalgo2009,Hamann04,Chen17,bischetti+2022,maiolino+2004,Doyee22}.

The classification boundary between ``mini-BAL'' and ``NAL'' AGN in not sharp, with different authors giving different definitions in terms of width and/or blueshift of the absorption \citep{Elvis00,Laha21,rodriguez-hidalgo2009,Hamann04,Chen17,Doyee22}. Here simply give a generic classification as mini-BAL/NAL without aiming at a more specific category.

It should be noted that some past works have reported some rare starburst galaxies showing outflows with high velocities, even in excess of 1000~km/s 
\citep{Diamond-Stanic12,Diamond-Stanic21,Sell14,Perrotta21}. However, these outflows are traced by lower ionization transitions (MgII absorption and [OIII]  emission). More importantly, a close inspection of those cases reveal that each of them show some AGN signature ([NeV] emission and/or broad MgII emission and/or broad H$\beta$ emission and/or X-ray emission and/or located in the AGN or composite region of diagnostic diagrams); therefore, although the AGN contribution to the bolometric luminosity of these galaxies may be arguable (also taking into account the variable nature of AGN), it is likely that the high velocity outflows seen in these rare cases is actually driven by the AGN that they host.

Finally, it should be noted that the CIV absorption trough goes nearly to zero (as in many BAL QSOs), which would imply total covering factor of the emitting source by the outflowing ionized gas along our line of sight. However, the errors 
leave scope for a contribution of 30\% of the emission potentially not covered by the CIV absorption, which can be associated with the extended host galaxy. Yet, if higher S/N data confirms the CIV trough going to zero, this would imply that the outflow has an extent covering also the host galaxy, i.e. $\sim$400 pc, which would be fully consistent with recent findings of BAL outflows extending on scales of up to several kpc \citep{arav+2018, xu+2019, byun+2022, walker+2022, Arav18}.

Given that CIV is a resonant line, the observed redshifted emission is also tracing the CIV counterpart of the redshifted Ly$\alpha$ emission seen in \cite{bunker_jades_2023}. i.e. the receiding side of the outflow.

We finally note that the spectrum between Ly$\alpha$ and the NV doublet show the tenative signature of a NV blueshifted trough (Fig.\ref{fig:spectra_appendix}d), which would be associated with the highly ionized outflow, but it requires additional data to be confirmed.

\subsection{Constraints from other emission lines and diagnostics}

Although the paper focuses on a few lines discussed in the main text, in this section we also discuss other emission lines that have either lower S/N, more severe blending, or whose upper (or lower) limits give line ratios that are fully consistent with the AGN scenario.

\subsubsection{MgII and CIII]}

The MgII2796,2804 doublet is well resolved with the grating and in principle a good tracer of gas density in the range between 10$^9$ and 10$^{14}$ cm$^{-3}$.
However, the observed ratio, $1.36^{+0.67}_{-0.42}$, is so uncertain to be consistent both with the low density regime (ratio of $\sim$1) and the high density regime (ratio of $\sim$2). Moreover, even if additional data allows constraining the MgII doublet ratio more tightly, these are resonant transitions, which are therefore strongly sensitive to the optical depth and radiative transfer effects \citep{Chisholm20}.

The CIII]$\lambda\lambda$1907,1909 doublet would also be a good density tracer, as the ratio of its two components is primarily sensitive to the gas density and changes strongly between $10^4$ and $10^6$~cm$^{-3}$ (with the blue component $\lambda$1907 going to zero at high densities), similar to the NIV] doublet. However, as discussed above, the two components are unresolved with the grating, and we cannot obtain reliable constraints on the gas density nor on the line widths. Even more importantly,
CIII] emission is commonly seen also in star forming galaxies and in the NLR of AGN, so it may partially come also from the low density ISM of the host galaxy, as it is the case for [NeIII].
 Indeed, as already mentioned, recent IFS observations show CIII] to be resolved on scales of several 100~pc \citep{Maiolino23b}. It is interesting that when fitted with narrow and broad components, as discussed in the previous section, the narrow component gives widths formally consistent with the [NeIII], while the broad component is consistent with the NIV width.

\subsubsection{NV and NeV}

Additional transitions from species requiring ionizing photon energy higher than about 60
eV, such as NV and NeV (in addition to NeIV seen in GN-z11), are often seen as evidence for the presence of an AGN.
Yet, conversely, their absence should not be necessarily seen as evidence for the absence
of an AGN, as often these lines are weak even in AGN and remain undetected if the S/N is
not high enough \citep{kuraszkiewicz+2004, nagao+2006b, cleri+2023}. Moreover, the intensity of these lines varies strongly from case to case.

\begin{figure}[h]%
\centering
\includegraphics[width=0.8\textwidth]{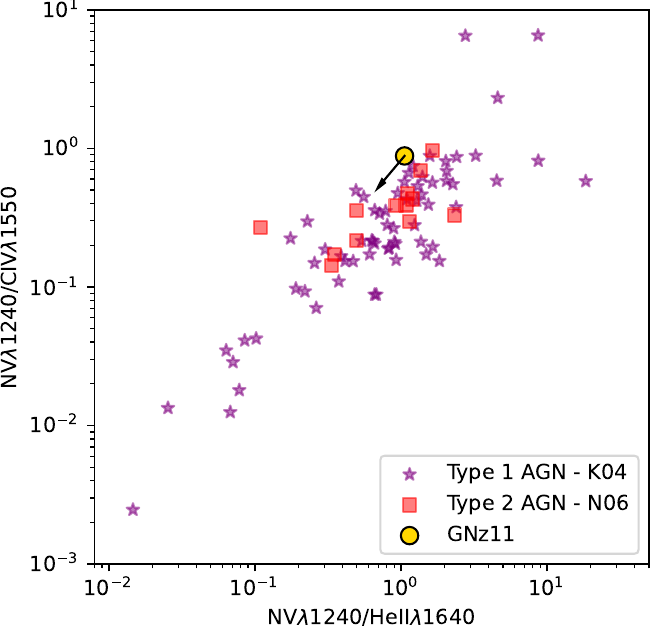}
\caption{NV/CIV versus NV/HeII flux ratio diagram for GN-z11 (golden circle) compared with the ratios observed for the broad lines of type 1 AGN (purple stars, \cite{kuraszkiewicz+2004}) and for the NLR of type 2 AGN (red squares, \cite{nagao+2006b}), illustrating that the non-detection of NV for GN-z11 is not constraining and consistent with the AGN scenario.} \label{fig:nv}
\end{figure}

With the prism it is not possible to assess the presence of NV because it is blended with Lya and its damping wing. Regarding the gratings, the G140M band, where NV is redshifted, is the least sensitive of the three medium resolution spectra. Although there is a hint of the NV doublet (Fig.\ref{fig:spectra_appendix}, a 2$\sigma$ integrated signal) we obviously do not quote it as a tentative detection.
The inferred upper limit on the NV emission is not very constraining, but the important aspect in the context of this paper is that it is still fully consistent with the presence of an AGN. We demonstrate this in Fig.\ref{fig:nv}, where the upper limits on the NV/CIV and NV/HeII ratios for GN-z11 are compared with a sample of the broad lines in type 1 AGN \citep{kuraszkiewicz+2004} and also with a sample of the NLR in type 2 AGN \citep{nagao+2006b}, and clearly illustrating that the non-detection of NV is fully consistent with the AGN scenario. 

It is interesting also to compare NV with NIV, as this ratio is not dependent on the nitrogen abundance, although NIV is detected (or reported) less frequently in AGN. In the well studied type 1.8 AGN at z=5.5, GS-3073 \citep{vanzella+2010, grazian+2020, ubler+2023}, the NV is five times fainter than NIV, which would be totally undetected in our spectrum.
In the type 1 quasars explored by \cite{Glikman07} the NIV broad line is very strong, while NV is undetected, with an upper limit that is about ten times lower than the NIV flux.

NeV is also not detected, neither in the grating nor in the prism spectrum. The upper limit on the NeV/NeIII ratio is about 0.2. However, \cite{cleri+2023} has shown that AGN models can have NeV/NeIII as low as $10^{-2}-10^{-4}$. Hence also the non-detection of NeV is not constraining about the presence of an AGN.

Finally, we note that AGN accreting at super-Eddington rates have a lower energy cutoff, hence are less likely to emit hard  photons that can produce highly ionised species, such as NV and NeV.

\subsubsection{HeII and CIV}

HeII is detected in the prism and, more marginally, in the grating (Fig.\ref{fig:spectra_appendix}).

As already discussed, CIV is clearly detected in the grating, but with a
P-Cygni profile, hence its flux is a lower limit because of self-absorption.

\begin{figure}[h]%
\centering
\includegraphics[width=0.99\textwidth]{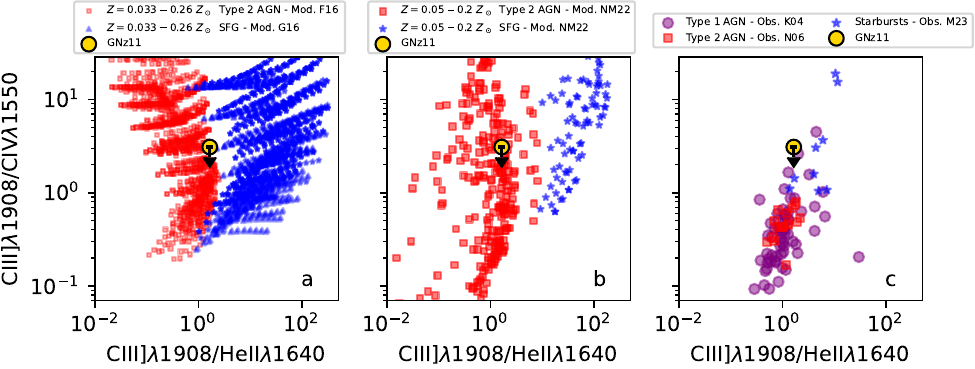}
\caption{CIII]/CIV versus CIII]/HeII flux ratio diagrams for GN-z11 (golden circle) compared with: a)
the AGN-NLR models (red squares) by \cite{Feltre16} and SF galaxies (blue stars) by \cite{Gutkin16} (left) and with b) the AGN-NLR models (red squares) and SF models (blue triangles) by \cite{Nakajima22} (centre). All models have been chosen in a low metallicity range, around the value inferred by \cite{bunker_jades_2023} for GN-z11. c) Comparison of GN-z11 with the
 ratios {\it observed} for the broad lines of type 1 AGN (purple circles, \cite{kuraszkiewicz+2004}), narrow lines of type 2 AGN (red squares, \cite{nagao+2006b}), and starburst galaxies (blue stars, \cite{mingozzi_classy_2022})}. \label{fig:civ_heii}
\end{figure}

The interpretation of these limits using photoionization models
is very much model-dependent. We illustrate this in
Fig.~\ref{fig:civ_heii}a-b. 
Specifically, Fig.~\ref{fig:civ_heii}a, as in \cite{bunker_jades_2023}, shows the location of GN-z11 on the  CIII]/CIV versus
HeII/CIII] diagram and where the red-squared and blue-starred symbols show the
location of models from \cite{Feltre16} and \cite{Gutkin16} for the the NLR of AGNs and for
star forming galaxies, respectively, and in a range of about $\pm$0.3 (see legend)
dex of the metallicity inferred by \cite{bunker_jades_2023} for GN-z11. Clearly, GN-z11 can be
consistent with both AGN and star forming models.

Fig.\ref{fig:civ_heii}b shows the same diagram where we instead plot the
models from \cite{Nakajima22}, in the same (low) metallicity range for both AGN and SF galaxies. Clearly,
in this case GN-z11 is much more consistent with the AGN models and inconsistent
with the models for star forming galaxies.

Yet, if the permitted and semi-forbidden lines of GN-z11 are coming
from the BLR, as argued in this paper, then neither of the models above
actually apply, as they are developed for the low density environments of the
NLR and HII regions. It is therefore more instructive to compare with the line
ratios {\it
observed} in the BLR of type 1 AGN. These are taken from the compilation of
\cite{kuraszkiewicz+2004} and shown with purple-circles in Fig.~\ref{fig:civ_heii}c. Clearly, the
line ratios observed in GN-z11 are fully consistent with the broad lines of type
1 AGN. For completeness, in the same panel we also plot the ratios observed for the NLR of type 2 AGN, compiled by \cite{nagao+2006b} (mostly overlapping with the ratios observed for the broad lines),  and the star forming galaxies from the CLASSY survey \cite{mingozzi_classy_2022}.

 \subsection{The WR scenario}

In this section we discuss the scenario recently proposed that GN-z11 may be similar to local WR galaxies  \cite{Senchya23}.

The HeII marginal detection shows a potentially broad profile ($\sim$ 1000 km/s, although the broad wings are mostly in the noise), which may be associated with the inner BLR, but also may resemble the broad HeII profile characteristic of WR stars. Therefore, there might be a contribution from WR stars and, specifically, WN stars, given the strong nitrogen lines.

However, there are various spectral features that cannot be accounted for in the WN scenario.

WN stars  are also characterised by very strong NIV$\lambda$1718 resonant emission, stronger than the NIV$\lambda$1486, and typically with a prominent P-Cygni profile \citep{Hainich14}. In GN-z11, despite the very strong NIV$\lambda$1486, there is no trace of the NIV$\lambda$1718 line. Fig.\ref{fig:spectra}f shows the spectrum of GN-z11 at the expected location of NIV$\lambda$1718 and where the shaded red region shows the expected intensity of the line, based on the strength of the NIV$\lambda$1486 line. Clearly, the GN-z11 spectrum is totally inconsitent with the presence of the NIV$\lambda$1718 WR signature.

Additionally, neither [NeIV]$\lambda$2424 nor CII$^*\lambda$1335 are ever seen associated with the WR population \citep{Hainich14}.

Finally, even if WN show prominent NIII] emission, the strength of the $\lambda$1754 component of the multiplet is much fainter in WR galaxies such as Mrk966 \citep{berg+2022}, and consistent with densities typical of the ISM.

Summarising, although WR stars might be present in GN-z11, they are unlikely to dominate the excitation of most nebular lines.

Table \ref{tab:scenarios} summarises more schematically the 
observational features consistent or inconsistent with the AGN scenario, the WR scenario, and a compact starburst without WR stars.

\begin{table}
    \centering
    \renewcommand{\arraystretch}{1.5}
    \caption{Features observed in GN-z11 versus possible energy sources.}
    
    \begin{tabular}{p{0.3\linewidth} | p{0.18\linewidth}  p{0.18\linewidth}  p{0.18\linewidth}}
    \hline
     & \multicolumn{3}{c}{\bf Source} \\
     \hline 
     & {\bf AGN}  & {\bf Wolf-Rayets} & {\bf Nuclear }  \\
     {\bf Feature} &  {\bf (NLSy1)} &  & {\bf Starburst}  \\
        &  &  & {\bf (without WR)} \\
    \hline
    Gas density $\rm >10^9~cm^{-3}$     & Consistent & Inconsistent & Inconsistent \\
    (from NIII \& NIV lines) & &  &  \\
    NeIV detection & Consistent  & Inconsistent & Inconsistent \\
    CII$^*$ detection  & Consistent  & Inconsistent & Inconsistent \\
    HeII broad & Consistent  & Consistent & Inconsistent \\
    NeIV(1718)/NeIV(1485)$<$10 & Consistent  & Inconsistent & Consistent \\
    $\rm V_{wind}>800 km/s$ & Consistent  & Inconsistent & Inconsistent \\
    UV-opt. cont. slope = 2.3 & Consistent  & Consistent & Consistent \\
    X-ray upper limit & Consistent  & Consistent & Consistent \\
    CIII]/CIV vs CIII]/HeII & Consistent  & Consistent & Consistent \\
    NV upper limit & Consistent  & Consistent & Consistent \\
    \hline
    \end{tabular}
    \label{tab:scenarios}
\end{table}

\subsection{Photoionization modelling}

We used the \textsc{Cloudy} photoionization code \citep{Ferland17} to explore the effect of varying physical conditions on some emission line ratios constrained by JWST/NIRSpec. The primary goal is to explore the ratios of emission lines within a given doublet or multiplet, hence lines of the same ion (specifically NIII and NIV) that are effectively insensitive to the chemical abundance and ionization parameter, while sensitive to density and only with secondary dependence on temperature. For this reason, the details of the photoionization models are not as critical as when exploring other line ratios. We considered a nebula of constant pressure in plane-parallel geometry. However, we have verified that other scenarios, such as a cloud with constant density, do not impact our findings. For completeness, we considered both AGN and stellar templates for the shape of the incident radiation field. Its normalization is set by the ionization parameter, defined as $U \equiv \Phi_\text{H} / (n_\text{H} c)$ where $\Phi_\text{H}$ is the surface flux of hydrogen-ionizing photons at the illuminated face of the nebula, $n_\text{H}$ is the number density of hydrogen, and $c$ is the speed of light. The hydrogen density and ionization parameter were varied in logarithmic steps of $1$, respectively from $n_\text{H} = 1 \, \mathrm{cm^{-3}}$ up to $n_\text{H} = 10^{14} \, \mathrm{cm^{-3}}$, and starting at $\log_{10} U = -3$ and ending at $\log_{10} U = -1$ \citep[e.g.][]{Feltre16, Gutkin16, Nakajima22}.

In the AGN scenario, we adopted the multi-component continuum template implemented in \textsc{Cloudy}, consisting of a black body and a power-law, varying the black body temperature ($T_\text{AGN} =10^6~K$ and $10^7~K$) while fixing the power-law slope to $\alpha = -1.4$ (note that this is the slope undelying the black body at energies above the Ly-edge) and leaving other optional parameters as default. For the AGN models, we considered gas-phase metallicities of $Z_\text{neb} = 0.1 \, \mathrm{Z_\odot}$ and $Z_\text{neb} = 1 \, \mathrm{Z_\odot}$. The star-forming models, on the other hand, are restricted to $Z_\text{neb} = 0.1 \, \mathrm{Z_\odot}$, since the hard ionising spectra of metal-poor stars are essential to form sufficient triply ionised nitrogen (requiring $47.5 \, \mathrm{eV}$), whose presence in GN-z11 is evidenced by the strong NIV emission ($\text{EW}_\mathrm{NIV \, 1486} = 9.0 \pm 1.1 \, \text{\AA}$; \cite{bunker_jades_2023}), while metal rich stars would not produce enough hard ionizing photons to make the NIV line visible. In the star-formation scenario, we employed stellar population synthesis models including binary stars generated by \textsc{bpass} v2.1 \citep{Eldridge17} for a single burst of star formation (with varying ages, $t_* / \mathrm{Myr} \in \{ 1, 10, 100 \}$), assuming the same metallicity as the gas (i.e. $Z_* = Z_\text{neb} = 0.1 \, \mathrm{Z_\odot}$) and a \cite{Salpeter55} initial mass function (IMF) that ranges in stellar mass from $1 \, \mathrm{M_\odot}$ to $100 \, \mathrm{M_\odot}$. Both in the AGN and star-formation cases, calculations are run until a neutral hydrogen column density of $N_\text{HI} = 10^{21} \, \mathrm{cm^{-2}}$ is reached to ensure that in all models the nebula is matter bounded; we note, however, that the highly ionised nitrogen lines are produced in the very inner part of the cloud, such that the boundary conditions do not significantly affect our results. In total, this results in a parameter grid of $15$ different densities, $3$ ionization parameters, $3$ temperatures or stellar ages, $2$ or $1$ metallicities for the AGN and star-formation models respectively, or a total of $15 \times 3 \times (3 \times 2 + 3 \times 1) = 405$ possible model configurations.

The relevant nitrogen line ratios for all of these (except for $8$ cases where \textsc{Cloudy} reported a failure) are shown in Fig.~\ref{fig:cloudy}, from which we conclude that they are consistent between the AGN and star-formation scenario, and their density dependence is largely independent of ionization parameter, metallicity, or the precise shape of the incident radiation field (i.e. AGN or star formation and the corresponding parameter $T_\text{AGN}$ or $t_*$). 

At high densities, the NIV$\lambda$1483/NIV$\lambda$1486 ratio approaches zero ($n_\text{H} \gtrsim 10^6 \, \mathrm{cm^{-3}}$), while NIII]$\lambda$1754 plateaus at a fractional contribution to the multiplet of $\sim 0.23$ at higher densities still ($n_\text{H} \gtrsim 10^{10} \, \mathrm{cm^{-3}}$), both pointing towards the presence of a broad line region in GN-z11 given the observed values.

\begin{figure}[h]%
\centering
\includegraphics[width=0.95\textwidth]{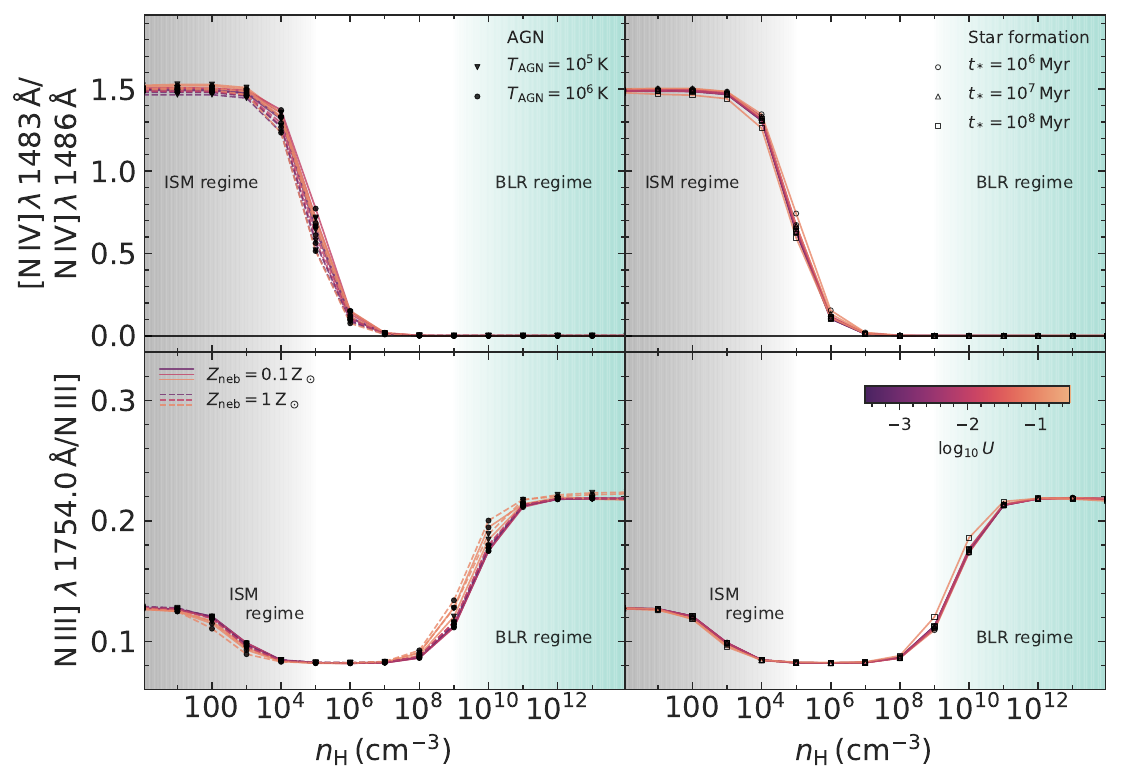}
\caption{Same as Fig.\ref{fig:cloudy} but where we have separated the photoionization models for AGN (left) and Star Forming galaxies (right).} \label{fig:cloudy_two_panels}
\end{figure}

 Finally, to increase the readibility of Fig.\ref{fig:cloudy} we have separated the AGN and Star Forming models in two separate panels in Fig.\ref{fig:cloudy_two_panels}.

\subsection{Continuum shape}

If GN-z11 is a type 1 AGN, then we should be directly seeing the light from the accretion disc.
In the case that the accretion disc dominates, the UV-to-optical continuum should
follow a simple power-law in the form $\rm F_{\lambda} \propto \lambda ^{\beta}$ with $\beta=-7/3\approx-2.33$ \citep{shakura+sunyaev1973}, as indeed observed in type 1 AGN, and NLSy1  \citep{capellupo+2015, leighly+moore2004}, modulo the UV turnover whose wavelength increases with black hole mass and also modulo effects of dust reddening, which often makes the spectrum redder.

\begin{figure}[h]%
\centering
\includegraphics[width=0.95\textwidth]{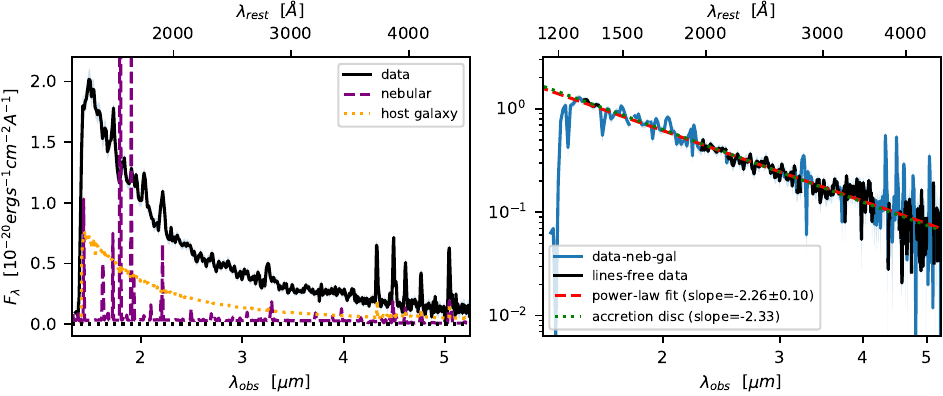}
\caption{Left: Observed prism spectrum of GN-z11 (black solid) compared with the (maximum) contribution from the host galaxy of the AGN as inferred by \cite{tacchella_jades_2023} (orange dotted), and the nebular emission inferred from a simple Cloudy model (purple dashed) normalized to the H$\gamma$ flux not included in the galaxy model.
Right: GN-z11 spectrum subtracted of the galactic and nebular continua, in a log-log scale, whose regions not affected by emission lines (solid black) have been fitted with a simple powerlaw (red-dashed) resulting into a slope of $-2.26\pm 0.10$, consistent with the slope expected for an accretion disc ($-2.33$, dotted green line).} \label{fig:cont_dec}
\end{figure}

In the case of GN-z11 the spectrum is contributed to also by the underlying galaxy identified by \cite{tacchella_jades_2023} in the NIRCam images. This component is significantly fainter than the nuclear point-like component. It is difficult to quantitatively establish its contribution to the spectrum, because part of the light may fall outside the shutter, and in a different fraction in the four dither/pointing positions, and not easy to reconstruct because of the slight positional uncertainties discussed above. In Fig.\ref{fig:cont_dec}a we show the contribution from the galactic component (dotted-orange line) to the spectrum, assuming that the entire light of the galaxy is captured by the spectrum, corresponding to about 1/3 of the flux, and using the spectral template inferred by \cite{tacchella_jades_2023} for the extended component.

The additional component to take into account is the nebular continuum associated with the BLR (as well as any other ionized gas in the host galaxy). The BLR typically has a low covering factor \citep{Maiolino01a}, therefore the nebular continuum is not expected to be strong, but its contribution must be quantified. In most physical conditions typical of the ionized gas in the BLR, NLR or HII regions, the nebular continuum is linked to the intensity of the Balmer lines. We have estimated the nebular emission using a Cloudy model with metallicity of 0.1~Z$_\odot$ and a density of $10^6$~cm$^{-3}$ (between the BLR and ISM origin scenarios) and normalized to have the same H$\gamma$ flux as observed in the spectrum of GN-z11. The nebular spectrum does not change drastically as a function of density, except obviously for the emission of the forbidden and semi-forbidden lines; however, our focus is on the nebular continuum, so we ignore the mismatch of the emission lines, as a detailed photoionization modelling of their flux is beyond the scope of this paper. We note that the nebular continuum is also included in the model spectrum fit to the extended component by \cite{tacchella_jades_2023}; therefore, not to include it twice, we have measured the H$\gamma$ flux in the \cite{tacchella_jades_2023} spectrum and normalized the Cloudy nebular spectrum only to the H$\gamma$ flux obtained by the difference of the observed value and the flux in the \cite{tacchella_jades_2023} model spectrum. The resulting nebular spectrum is shown with a dashed purple line in Fig.\ref{fig:cont_dec}a; once again, the mismatch of the emission lines should be disregarded, as the goal is not to reproduce them with the Cloudy model.

Fig.\ref{fig:cont_dec}b shows again the observed spectrum, in log-log scale, where the galactic and nebular components have been subtracted. Although the noise is large, especially at long wavelengths, also as a consequence of the model subtraction procedure,  the resulting spectrum is well fitted by a simple powerlaw, in the parts not affected by emission lines.
The best-fitting slope is --2.26$\pm$0.10, hence consistent with the continuum expected from an accretion disc.
Note that this is not evidence in support of the presence of an AGN,  as also young galaxies may have power-law shapes, it is only meant to show consistency with the AGN scenario.

We finally note that, although with large scatter, the UV spectrum of AGN often shows a FeII hump between $\sim$2300\AA \ and $\sim$3100\AA \ \citep{Maiolino03,DeRosa11,Mazzucchelli17,Shin19,Sameshima20}. The prism spectrum of GN-z11 does not show an obvious FeII bump, although a more detailed analysis and modelling is required to assess the presence or absence of such a bump, which is deferred to a separate paper. However, we note that at such early epochs there is little time for the ISM to be enriched with iron through the SNIa channel \citep{Maiolino19}, so a weak or absent FeII bump would not be unexpected.

\subsection{Variability}

The luminosity of AGN can be variable, from a few percent to a factor of a few, on short (days) and long (years) timescales. We have investigated the possible presence of variability. Before the recent NIRCam images obtained in February 2023 \citep{tacchella_jades_2023}, deep photometric observations were obtained with HST about ten years earlier \citep{oesch+2014, oesch_remarkably_2016}, corresponding to about one year in the rest frame of GN-z11. Most of the HST photometric datapoints have errorbars that are too large to be useful for constraining variability. However, the photometric point reported in 
\cite{oesch+2014} with the F160W filter has a relatively well constrained value: 150$\pm$10~nJy, within an aperture of 0.35$''$.
NIRCam does not have the same filter, however, the photometry obtained in the F150W filter can be used and transposed to the  F160W filter by using the NIRSpec prism spectrum. After extracting photometry from a 0.35$''$ aperture (as in \cite{oesch+2014}), and extrapolating with the NIRSpec spectrum, we obtain a F160W equivalent photometry of 141$\pm$2~nJy, which is consistent with the HST previous photometry within 1$\sigma$. If we consider that about 30\% of the flux is diluted by the host galaxy, the comparison of the photometry between the two epochs would indicate a variability of 10\% at only 1$\sigma$. This is certainly not a detection of variability, but it is consistent with the range of variability amplitudes observed in NLSy1 and, more broadly, in type 1 AGN \citep{Ai13}.

\subsection{X-ray emission}

GN-z11 is not detected in X-rays. Flux limits are obtained from the Chandra Deep Field North which was a 2 Ms observation performed in 2002 (see \cite{Xue16} for final results). Their sensitivity map gives a point source limit in the Soft (0.5-2 keV), Hard (2-7 keV) and Full (0.5-7 keV) bands of $1.54\times 10^{-17}$, $7.9\times 10^{-17}$ and $4.9\times 10^{-17}$ $\rm erg ~cm^{-2}~s^{-1}$. Source detection requires a no-source probability P$<$0.004. The tightest limit in the Soft Band translates to a restframe 5.8-23.2 keV luminosity limit at z=10.6 of 2.2$\times 10^{43}~\rm erg~s^{-1}$.   Assuming a typical NLS1 photon index of 2.3 means that $\rm L_{\rm X}$(2-10 keV) is less than $3\times 10^{43}~\rm erg~s^{-1}$.

The Bolometric Correction for NLS1 in the 2-10 keV band, $\rm BC_{\rm X}$, is about 100 \citep{Vasudevan07}. There is a significant systematic uncertainty here due to the unseen flux in the FUV where the emission is expected to peak (see Fig.3 in \citep{Buisson18}). Moreover the 2-10 keV flux entirely orginates from the corona, the early development of which and possible dependence on black hole spin are unknown (\cite{Netzer19} cautions against using his X-ray BC values for NLS1).  Proceeding with $\rm BC_{\rm X}=100$ means that the Chandra upper  limit is almost 3 times above the luminosity inferred from the JWST flux at 1400\AA.  We predict a conservative SB flux of $5\times  10^{-18}~ \rm erg~cm^{-2}~s^{-1}$. This would be detectable in about 1 Ms with the candidate NASA Probe mission AXIS.  The coronal emission from local NLS1s is highly variable and the above BC represents a mean value (note that the intrinsic disc flux seen in the UV is much less variable, e,g, \cite{Buisson18}).

\subsection{BH mass estimate}

For the vast majority of high redshift AGN the black hole masses are inferred using single-epoch measurements and the so-called virial relations, i.e. relations between the black hole mass, the width of the lines of the BLR and the continuum or line luminosity \citep[e.g.][]{Willott10, Trakhtenbrot11, Shen11, Shen19, Pensabene20, izumi_subaru_2021, Harikane23, Mezcua23}. These relations are calibrated on nearby AGN, using either reverberation mapping techniques and/or direct dynamical measurements of the black hole. The black hole mass scales about as the square power of the width of the BLR lines and about as the square root power of the luminosity, with a proportionality constant that depends on the specific waveband (or line) for the luminosity estimation.

The most accurate virial relations would be those using H$\alpha$ and H$\beta$. In our case H$\gamma$ could be used as a proxy. However, as discussed, the Balmer lines are likely contributed to by the star formation in the host galaxy, hence not reliable to trace the black hole mass.

The CIII] doublet is also sometimes used to infer the black hole mass. However, this is not well resolved and, as for the case of the Balmer lines, this is likely contaminated by the ISM and star formation in the host galaxy.

MgII is often used. In our case the MgII doublet is clearly detected, but the S/N is fairly low for the measurement of the width (Fig.\ref{fig:spectra_appendix}). If we take the width resulting from the fit and the relation provided by \cite{Vestergaard09}:
$$\rm
\log{\left( \frac{M_{BH}}{M_\odot} \right)} =
  6.86 + 0.5 
  \log{\left( \frac{(\lambda L_{\lambda})_{3000\text{\AA}}}{10^{44}erg/s} \right)} +
  2 \log{\left( \frac{FWHM_{MgII}}{10^{3}kms/s} \right)}
$$

then we get a black hole mass of $\rm 1.4\times 10^6~M_\odot$. However, given the low S/N on the MgII doublet, we prefer to use as representative width of the BLR lines the profile of the high S/N and isolated NIV line. If we adopt this width into the equation above, we obtain a black hole mass of $\rm 1.6 \times 10^6~M_\odot$.
The uncertainty is totally dominated by the scatter in the virial scaling relation, which is about 0.3~dex \citep{Kormendy13}.

Additionally,  there are various other systematic uncertainties and caveats that can affect the black hole mass estimate. To begin with, it is not obvious that the local virial relations apply at high redshift. The main issue is whether the dependence of the BLR radius on luminosity, evolves with reshift or not. The most plausible scenario is that the square root dependence of the BLR radius from luminosity is primarily set by the dust sublimation radius. \cite{Maiolino23c} argue that, given the extremely high densities in the nuclear region of AGN (hence high optical thickness even at very low dust-to-gas ratios), unless the nuclear region is totally devoid of dust,  the same $\rm R_{BLR}-L$ relation is unlikely to evolve with redshift.
More problematic is assessing whether the virial relations depend on the accretion rate or not. On the one hand, \cite{Marconi2008} argue that the effect of radiation pressure is to reduce the effective gravitational force on the clouds of the BLR; the net result is that the standard virial relations applied to BHs accreting close to the Eddington rate could underestimate the BH mass by a factor of several. On the other hand, reverberation mapping of AGN accreting at super-Eddington have revealed that in these cases the size of the BLR is a factor of several, and up to an order of magnitude, smaller than expected from the $\rm R_{BLR}-L$ relation for sub-Eddington AGN (\cite{Khatu23} and references therein), which would imply that the standard virial relations over-estimate, by a factor of several, the BH masses in AGN accreting at super-Eddington. Overall, it is possible that the radiation pressure effect and the offset from the $\rm R_{BLR}-L$ relation might cancel each other out. However, currently it is not really possible to provide an accurate assessment on how much AGN accreting at or beyond the Eddintgon rate might deviate from the standard virial relations. 

Finally, the BH masses from other JWST studies at z$\sim$4--8
\citep{Kocevski23,Harikane23,ubler+2023,Ding22,Maiolino23c} are shown in Fig.\ref{fig:mbh_mst}. These are based on the H$\alpha$ or H$\beta$ width and flux. We clarify that 
these are re-estimated by using the same calibrations used in \cite{Reines15} for local galaxies.

\subsection{AGN bolometric luminosity estimate}

We derive the bolometric luminosity of the AGN by using the continuum luminosity at $\lambda _{rest}=1400$\AA \ and the luminosity-dependent bolometric correction given by \cite{Netzer19}:
$$ \rm
\frac{L_{bol}}{(\lambda L_{\lambda})_{1400\text{\AA}}} = 7~
\left( \frac{(\lambda L_{\lambda})_{1400\text{\AA}}}{10^{42}erg/s} 
\right) ^{-0.1}
$$
We also assume, as discussed in the previous sections, that 30\% of the continuum flux at this wavelength is due to the underlying galactic component (\citep{tacchella_jades_2023}) and that, therefore, the AGN continuum luminosity at this wavelength is 0.7 of the observed value.
We infer a bolometric luminosity of $1.08\times10^{45}$~erg/s. The resulting ratio between bolometric and Eddington luminosity is 5.5, also affected by an uncertainty of a factor of at least 2, coming from the uncertainty on the black hole mass.

\subsection{Comparison with cosmological and hydrodynamical simulations}

There is a vast literature discussing the formation of early black holes and on how they evolve in the first billion year, by using  hydrodynamical and cosmological simulations, as well as semi-analytical models. The production and elaboration of models in this area have recently seen surge with the goal of specifically interpreting the results from JWST. It is beyond the scope of this paper to provide an exhaustive description of the assumptions and results of the several models and simulations. However, in this section we briefly discuss that many of them can explain the properties of GN-z11, and provide some possible constraints on the seeding scenarios.

We start by considering the results obtained by \citep{Bennett23} from the FABLE hydrodynamical, cosmological simulation, in which they focused on largest halo at z=6 (with a virial mass $\rm M_{200} = 6.9\times 10^{12}~M_\odot$ of the Millennium box). The latter may appear an extreme choice, however we note that GN-z11 does live in an overdense region and likely at the core of a protocluster \citep{tacchella_jades_2023,Schotlz23}. In the FABLE simulation the BH seed has a mass of $\rm 10^5~M_\odot$ at z=13.  The accretion rate is capped to Eddington and uses the  
Bondi-Hoyle-Littleton-based formalism; however, as small scale, non-isotropic accretion is unresolved in the simulation, FABLE, like Illustris, uses a Bondi-Hoyle-Littleton rate boosted by a factor of 100. Feedback energy in FABLE scales as 10 per cent of the available accretion energy, $\dot{E} = \epsilon \dot{M} c^2$, where $\epsilon = 0.1$ is the radiative efficiency of the accretion flow. At high redshifts, this is primarily injected as thermal energy in the vicinity of the black hole, with a duty cycle of 25 Myr.
We overplot the fiducial model in \citep{Bennett23} in Fig.\ref{fig:mbh_z} (orange solid line, labelled as B23), illustrating that this can easily reproduce the mass of the BH in GN-z11
at z=10.6.  

In order to explain the most massive BHs observed at z$\sim6-7$, \citep{Bennett23} also explore the scenario of earlier seeding (z=18) and allowing the BH to accrete at up to two times the Eddington limit; in this case the model could explain a BH nearly 5 times more massive than GN-z11 at z=10.6.

\cite{Zhang_Trinity_23_IV} explored the early evolution of black holes using the TRINITY cosmological empirical model \cite{Zhang_Trinity_23_I}, which is based on halo statistics from N-body simulations and incorporating empirical galactic scaling relations. They specifically explore the case of GN-z11. They illustrate that its mass and BH to stellar mass ratio can be explained by their model starting with an intermediate mass seed of a few times 10$^3$ seeded at z=15, accreting on average at sub-Eddington rates, but  intermittently also at super-Eddington. Their track is shown with a solid-teal line in Fig.\ref{fig:mbh_z} (labelled as Z23).

Recently, \citep{Schneider23} have explored the properties of GN-z11 within the context of the semi-analytical model CAT. They find that the BH mass of GN-z11 and its location on the $\rm M_{BH}-M_{star}$ diagram can be interpreted both in terms of light seeds (at z=20-23) that can have super-Eddington accretion phases, or Eddington-limited heavy seeds formed at z=14-16. Out of their various tracks, Fig.\ref{fig:mbh_z} shows only two samples of their tracks, in the case of a light (red-solid) and a heavy seed (red-dashed), which can both reproduce the mass of GN-z11 at z=10.6 (labelled as S23). In both cases the semi-analytical model can also reproduce the black hole to stellar mass observed in GN-z11.

\cite{Volonteri23} have suggested that the detectability of accreting BHs at high redshift by JWST implies that these are likely originating from heavy seeds. Specifically, their models can reproduce the mass of GN-z11 at z=10.6 but only with seeds that are several times $\rm 10^5~M_\odot$, already in place before z=14. GN-z11 would fall in this category and the tracks obtained by \cite{Volonteri23} would also explain the BH/stellar mass ratio observed in GN-z11.

Other studies have proposed other scenarios, envisanging different seeding mechanisms, at different redshifts, and with different assumptions about the accretion and merging rates, and which are capable of reproducing the BH mass of GN-z11 by z=10.6, and generally also its BH-to-stellar mass ratio
\citep{zhu+2020, bhowmick+2022, smidt+2018,Trinca22}.

Summarizing, the properties of the BH in GN-z11 can be  explained with different assumptions, which can be broadly grouped in heavy seeds accreting at sub-Eddington rates, or intermediate/light seeds experiencing super-Eddington phases and/or modelled with a boosted Bondi accretion.

More statistics on objects like GN-z11 is required to discriminate between different scenarios. For the time being GN-z11 remains the most luminous object at z$>$10 in all HST Deep fields (including CANDLES and Frontier Fields). Hopefully, JWST observations on larger areas (e.g. in Cosmos-WEB) will find more AGN at z$>$10 similar to GN-z11. For the time being, as discussed in the text, it is interesting to note that models and simulations were expecting a few accreting black holes with masses in the range $\rm 10^6-10^7 ~M_\odot$ at 10$<$z$<$11 in the JADES Medium-Deep survey in the GOODS fields \cite{trinca+2023,Volonteri23}. Therefore, the discovery of a $\rm 2\times 10^6~M_\odot$ black hole in GN-z11 is not unexpected, and a few more might be present (probably accreting at a lower rate) in the GOODS fields.

\subsection{GN-z11 and its large scale environment}

 We have shown that the high Nitrogen enrichment of GN-z11 is likely restricted to the BLR, whose small mass and compact size has probably undergone very rapid chemical enrichment, requiring only a few SNe.
 
 We note that the high chemical enrichment of GN-z11 is not in contrast with the recent claim of pristine gas in the halo of GN-z11
  \cite{Maiolino23b}. Indeed, these claims are on totally different scales, with the pristine gas found several kpc away from GN-z11, while the high chemical enrichment is estimated to be in the nucleus of GN-z11. Regarding the claim of pristine gas in the halo of GN-z11,
  models expect that high-z massive galaxies may host pockets of pristine gas in their haloes, even down to z$\sim$3 \cite{Liu20,Venditti23}.\\

\bmhead{Acknowledgments}
This paper benefited from suggestions and discussions with several colleagues, in particular: G. Risaliti, A. Marconi, J. Bennett, S. Koudmani, D. Sijacki, R. Schneider, M. Volonteri, R. Valiante, A. Trinca, D. Berg, R. Ellis, J. Dunlop, M. Pettini, H. Katz, W.N. Brandt and R. Tripodi.
FDE, JS, LS, RM and TJL acknowledge support by the Science and Technology Facilities Council (STFC) and by the ERC through Advanced Grant 695671 "QUENCH". RM also acknowledges funding from a research professorship from the Royal Society.
AJB, GCJ and JC acknowledge funding from the "FirstGalaxies" Advanced Grant from the European Research Council (ERC) under the European Union’s Horizon 2020 research and innovation programme (Grant agreement No. 789056).
BER acknowledges support from the NIRCam Science Team contract to the University of Arizona, NAS5-02015. 
BRP, MP and SA acknowledge support from Grant PID2021-127718NB-I00 funded by the Spanish Ministry of Science and Innovation/State Agency of Research (MICIN/AEI/ 10.13039/501100011033). MP also acknowledges support from the Programa Atraccion de Talento de la Comunidad de Madrid via grant 2018-T2/TIC-11715.
CNAW, EE and FS acknowledge a JWST/NIRCam contract to the University of Arizona NAS5-02015.
DJE is supported as a Simons Investigator and by JWST/NIRCam contract to the University of Arizona, NAS5-02015.
ECL acknowledges support of an STFC Webb Fellowship (ST/W001438/1).
H{\"U} gratefully acknowledges support by the Isaac Newton Trust and by the Kavli Foundation through a Newton-Kavli Junior Fellowship.
JW acknowledges support from the ERC Advanced Grant 695671, ``QUENCH'', and the Foundation MERAC.
SC acknowledges support by European Union’s HE ERC Starting Grant No. 101040227 - WINGS.
The research of CCW is supported by NOIRLab, which is managed by the Association of Universities for Research in Astronomy (AURA) under a cooperative agreement with the National Science Foundation.
This research is supported in part by the Australian Research Council Centre of Excellence for All Sky Astrophysics in 3 Dimensions (ASTRO 3D), through project number CE170100013.

\bmhead{Author contributions}
RM, JS, JW, SC, FDE, AdG, HÜ, and ST contributed to the writing of the paper, methods and creation of figures. 
All authors have contributed to the interpretation of the results. 
NK, BRdP contributed to the design, construction and commissioning of NIRSpec. 
SAr, SCa, MC, JW, MP, and BRdP contributed to the NIRSpec data reduction and to the development of the NIRSpec pipeline. 
SAr contributed to the design and optimisation of the MSA configurations. 
AJB, CNAW, ECL, KB, and HÜ contributed to the selection, prioritisation and visual inspection of the targets. 
SCh, JC, ECL, RM, JW, FDE, TJL, MC, AdG, and LS contributed to analysis of the spectroscopic data, including redshift determination and spectral modelling. 
FDE, TJL, MC, BRdP, RM, SA, and JS contributed to the development of the tools for the spectroscopic data analysis, visualisation and fitting.
CW contributed to the design of the spectroscopic observations and MSA configurations. 
CNAW, CW, DJE, RM, and SAr contributed to the design of the JADES survey. 
EE, KNH, and CCW contributed to the design, construction, and commissioning of NIRCam. 
BER, DJE, IS, ST, CNAW, and ZJ contributed to the JADES imaging data reduction.
BER contributed to the JADES imaging data visualisation.

\bmhead{Data Availability}

The electronic version of the processed data used to produce the figures (including the 1D and 2D spectrum of GN-z11) is available at the JADAES web site
https://jades-survey.github.io/.

The NIRSpec raw data can be accessed at the JWST archive
http://archive.stsci.edu.

\bmhead{Author Information}

The authors declare that they have no competing financial interests. Correspondence and requests for materials should be addressed to R.M. (email: rm665@cam.ac.uk).

\bibliography{sn-bibliography,roberto}

\begin{thebibliography}{100}
\expandafter\ifx\csname url\endcsname\relax
  \def\url#1{\burl{#1}}\fi
\expandafter\ifx\csname urlprefix\endcsname\relax\def\urlprefix{URL }\fi
\providecommand{\bibinfo}[2]{#2}
\providecommand{\eprint}[2][]{\url{#2}}
\providecommand{\doi}[1]{\url{https://doi.org/#1}}
\bibcommenthead

\bibitem{inayoshi+2020}
\bibinfo{author}{{Inayoshi}, K.}, \bibinfo{author}{{Visbal}, E.} \&
  \bibinfo{author}{{Haiman}, Z.}
\newblock \bibinfo{title}{{The Assembly of the First Massive Black Holes}}.
\newblock \emph{\bibinfo{journal}{\araa}} \textbf{\bibinfo{volume}{58}},
  \bibinfo{pages}{27--97} (\bibinfo{year}{2020}) .

\bibitem{fan_quasars_2022}
\bibinfo{author}{Fan, X.}, \bibinfo{author}{Banados, E.} \&
  \bibinfo{author}{Simcoe, R.~A.}
\newblock \bibinfo{title}{Quasars and the intergalactic medium at cosmic dawn}.

\bibitem{volonteri+2023}
\bibinfo{author}{{Volonteri}, M.}, \bibinfo{author}{{Habouzit}, M.} \&
  \bibinfo{author}{{Colpi}, M.}
\newblock \bibinfo{title}{{What if young z > 9 JWST galaxies hosted massive
  black holes?}}
\newblock \emph{\bibinfo{journal}{\mnras}} \textbf{\bibinfo{volume}{521}}~(1),
  \bibinfo{pages}{241--250} (\bibinfo{year}{2023}) .

\bibitem{trinca_low-end_2022}
\bibinfo{author}{Trinca, A.} \emph{et~al.}
\newblock \bibinfo{title}{The low-end of the black hole mass function at cosmic
  dawn} \textbf{\bibinfo{volume}{511}}~(1), \bibinfo{pages}{616--640}
  (\bibinfo{year}{2022}) .

\bibitem{Banik19}
\bibinfo{author}{{Banik}, N.}, \bibinfo{author}{{Tan}, J.~C.} \&
  \bibinfo{author}{{Monaco}, P.}
\newblock \bibinfo{title}{{The formation of supermassive black holes from
  Population III.1 seeds. I. Cosmic formation histories and clustering
  properties}}.
\newblock \emph{\bibinfo{journal}{\mnras}} \textbf{\bibinfo{volume}{483}}~(3),
  \bibinfo{pages}{3592--3606} (\bibinfo{year}{2019}) .

\bibitem{Singh23}
\bibinfo{author}{{Singh}, J.}, \bibinfo{author}{{Monaco}, P.} \&
  \bibinfo{author}{{Tan}, J.~C.}
\newblock \bibinfo{title}{{The formation of supermassive black holes from
  Population III.1 seeds. II. Evolution to the local universe}}.
\newblock \emph{\bibinfo{journal}{\mnras}} \textbf{\bibinfo{volume}{525}}~(1),
  \bibinfo{pages}{969--982} (\bibinfo{year}{2023}) .

\bibitem{Bennett23}
\bibinfo{author}{{Bennett}, J.~S.}, \bibinfo{author}{{Sijacki}, D.},
  \bibinfo{author}{{Costa}, T.}, \bibinfo{author}{{Laporte}, N.} \&
  \bibinfo{author}{{Witten}, C.}
\newblock \bibinfo{title}{{The growth of the gargantuan black holes powering
  high-redshift quasars and their impact on the formation of early galaxies and
  protoclusters}}.
\newblock \emph{\bibinfo{journal}{\mnras}} \textbf{\bibinfo{volume}{527}}~(1),
  \bibinfo{pages}{1033--1054} (\bibinfo{year}{2024}) .

\bibitem{tacchella_jades_2023}
\bibinfo{author}{{Tacchella}, S.} \emph{et~al.}
\newblock \bibinfo{title}{{JWST NIRCam + NIRSpec: interstellar medium and
  stellar populations of young galaxies with rising star formation and evolving
  gas reservoirs}}.
\newblock \emph{\bibinfo{journal}{\mnras}} \textbf{\bibinfo{volume}{522}}~(4),
  \bibinfo{pages}{6236--6249} (\bibinfo{year}{2023}) .

\bibitem{bunker_jades_2023}
\bibinfo{author}{{Bunker}, A.~J.} \emph{et~al.}
\newblock \bibinfo{title}{{JADES NIRSpec Spectroscopy of GN-z11:
  Lyman-{\ensuremath{\alpha}} emission and possible enhanced nitrogen abundance
  in a z = 10.60 luminous galaxy}}.
\newblock \emph{\bibinfo{journal}{\aap}} \textbf{\bibinfo{volume}{677}},
  \bibinfo{pages}{A88} (\bibinfo{year}{2023}) .

\bibitem{Feltre16}
\bibinfo{author}{{Feltre}, A.}, \bibinfo{author}{{Charlot}, S.} \&
  \bibinfo{author}{{Gutkin}, J.}
\newblock \bibinfo{title}{{Nuclear activity versus star formation:
  emission-line diagnostics at ultraviolet and optical wavelengths}}.
\newblock \emph{\bibinfo{journal}{\mnras}} \textbf{\bibinfo{volume}{456}}~(3),
  \bibinfo{pages}{3354--3374} (\bibinfo{year}{2016}) .

\bibitem{Terao22}
\bibinfo{author}{{Terao}, K.} \emph{et~al.}
\newblock \bibinfo{title}{{Multiline Assessment of Narrow-line Regions in z 3
  Radio Galaxies}}.
\newblock \emph{\bibinfo{journal}{\apj}} \textbf{\bibinfo{volume}{929}}~(1),
  \bibinfo{pages}{51} (\bibinfo{year}{2022}) .

\bibitem{Lefevre19}
\bibinfo{author}{{Le F{\`e}vre}, O.} \emph{et~al.}
\newblock \bibinfo{title}{{The VIMOS Ultra-Deep Survey: evidence for AGN
  feedback in galaxies with CIII]-{\ensuremath{\lambda}}1908 {\r{A}} emission
  10.8 to 12.5 Gyr ago}}.
\newblock \emph{\bibinfo{journal}{\aap}} \textbf{\bibinfo{volume}{625}},
  \bibinfo{pages}{A51} (\bibinfo{year}{2019}) .

\bibitem{Hainich14}
\bibinfo{author}{{Hainich}, R.} \emph{et~al.}
\newblock \bibinfo{title}{{The Wolf-Rayet stars in the Large Magellanic Cloud.
  A comprehensive analysis of the WN class}}.
\newblock \emph{\bibinfo{journal}{\aap}} \textbf{\bibinfo{volume}{565}},
  \bibinfo{pages}{A27} (\bibinfo{year}{2014}) .

\bibitem{Vanden_Berk01}
\bibinfo{author}{{Vanden Berk}, D.~E.} \emph{et~al.}
\newblock \bibinfo{title}{{Composite Quasar Spectra from the Sloan Digital Sky
  Survey}}.
\newblock \emph{\bibinfo{journal}{\aj}} \textbf{\bibinfo{volume}{122}}~(2),
  \bibinfo{pages}{549--564} (\bibinfo{year}{2001}) .

\bibitem{Wu22}
\bibinfo{author}{{Wu}, Q.} \& \bibinfo{author}{{Shen}, Y.}
\newblock \bibinfo{title}{{A Catalog of Quasar Properties from Sloan Digital
  Sky Survey Data Release 16}}.
\newblock \emph{\bibinfo{journal}{\apjs}} \textbf{\bibinfo{volume}{263}}~(2),
  \bibinfo{pages}{42} (\bibinfo{year}{2022}) .

\bibitem{grazian+2020}
\bibinfo{author}{{Grazian}, A.} \emph{et~al.}
\newblock \bibinfo{title}{{On the AGN Nature of Two UV-bright Sources at
  z$_{spec}$ {\ensuremath{\sim}} 5.5 in the CANDELS Fields: An Update on the
  AGN Space Density at M$_{1450}$ {\ensuremath{\sim}} -22.5}}.
\newblock \emph{\bibinfo{journal}{\apj}} \textbf{\bibinfo{volume}{897}}~(1),
  \bibinfo{pages}{94} (\bibinfo{year}{2020}) .

\bibitem{berg+2022}
\bibinfo{author}{{Berg}, D.~A.} \emph{et~al.}
\newblock \bibinfo{title}{{The COS Legacy Archive Spectroscopy Survey (CLASSY)
  Treasury Atlas}}.
\newblock \emph{\bibinfo{journal}{\apjs}} \textbf{\bibinfo{volume}{261}}~(2),
  \bibinfo{pages}{31} (\bibinfo{year}{2022}) .

\bibitem{mingozzi_classy_2022}
\bibinfo{author}{Mingozzi, M.} \emph{et~al.}
\newblock \bibinfo{title}{{CLASSY} {IV}. exploring {UV} diagnostics of the
  interstellar medium in local high-z analogs at the dawn of the {JWST} era}.
\newblock \emph{\bibinfo{journal}{\apj}} \textbf{\bibinfo{volume}{939}},
  \bibinfo{pages}{110} (\bibinfo{year}{2022}) .

\bibitem{osterbrock+pogge1985}
\bibinfo{author}{{Osterbrock}, D.~E.} \& \bibinfo{author}{{Pogge}, R.~W.}
\newblock \bibinfo{title}{{The spectra of narrow-line Seyfert 1 galaxies.}}
\newblock \emph{\bibinfo{journal}{\apj}} \textbf{\bibinfo{volume}{297}},
  \bibinfo{pages}{166--176} (\bibinfo{year}{1985}) .

\bibitem{Mathur12}
\bibinfo{author}{{Mathur}, S.}, \bibinfo{author}{{Fields}, D.},
  \bibinfo{author}{{Peterson}, B.~M.} \& \bibinfo{author}{{Grupe}, D.}
\newblock \bibinfo{title}{{Supermassive Black Holes, Pseudobulges, and the
  Narrow-line Seyfert 1 Galaxies}}.
\newblock \emph{\bibinfo{journal}{\apj}} \textbf{\bibinfo{volume}{754}}~(2),
  \bibinfo{pages}{146} (\bibinfo{year}{2012}) .

\bibitem{leitherer+2011}
\bibinfo{author}{{Leitherer}, C.}, \bibinfo{author}{{Tremonti}, C.~A.},
  \bibinfo{author}{{Heckman}, T.~M.} \& \bibinfo{author}{{Calzetti}, D.}
\newblock \bibinfo{title}{{An Ultraviolet Spectroscopic Atlas of Local
  Starbursts and Star-forming Galaxies: The Legacy of FOS and GHRS}}.
\newblock \emph{\bibinfo{journal}{\aj}} \textbf{\bibinfo{volume}{141}}~(2),
  \bibinfo{pages}{37} (\bibinfo{year}{2011}) .

\bibitem{Du16}
\bibinfo{author}{{Du}, X.}, \bibinfo{author}{{Shapley}, A.~E.},
  \bibinfo{author}{{Martin}, C.~L.} \& \bibinfo{author}{{Coil}, A.~L.}
\newblock \bibinfo{title}{{The Kinematics of C IV in Star-forming Galaxies at z
  {\ensuremath{\approx}} 1.2}}.
\newblock \emph{\bibinfo{journal}{\apj}} \textbf{\bibinfo{volume}{829}}~(2),
  \bibinfo{pages}{64} (\bibinfo{year}{2016}) .

\bibitem{Elvis00}
\bibinfo{author}{{Elvis}, M.}
\newblock \bibinfo{title}{{A Structure for Quasars}}.
\newblock \emph{\bibinfo{journal}{\apj}} \textbf{\bibinfo{volume}{545}}~(1),
  \bibinfo{pages}{63--76} (\bibinfo{year}{2000}) .

\bibitem{Senchya23}
\bibinfo{author}{{Senchyna}, P.}, \bibinfo{author}{{Plat}, A.},
  \bibinfo{author}{{Stark}, D.~P.} \& \bibinfo{author}{{Rudie}, G.~C.}
\newblock \bibinfo{title}{{GN-z11 in context: possible signatures of globular
  cluster precursors at redshift 10}}.
\newblock \emph{\bibinfo{journal}{arXiv e-prints}}
  \bibinfo{pages}{arXiv:2303.04179} (\bibinfo{year}{2023}) .

\bibitem{Zhang_Trinity_23_IV}
\bibinfo{author}{{Zhang}, H.} \emph{et~al.}
\newblock \bibinfo{title}{{TRINITY IV: Predictions for Supermassive Black Holes
  at $z rsim 7$}}.
\newblock \emph{\bibinfo{journal}{arXiv e-prints}}
  \bibinfo{pages}{arXiv:2309.07210} (\bibinfo{year}{2023}) .

\bibitem{Schneider23}
\bibinfo{author}{{Schneider}, R.} \emph{et~al.}
\newblock \bibinfo{title}{{Are we surprised to find SMBHs with JWST at z
  {\ensuremath{\geq}} 9?}}
\newblock \emph{\bibinfo{journal}{\mnras}} \textbf{\bibinfo{volume}{526}}~(3),
  \bibinfo{pages}{3250--3261} (\bibinfo{year}{2023}) .

\bibitem{Volonteri23}
\bibinfo{author}{{Volonteri}, M.}, \bibinfo{author}{{Habouzit}, M.} \&
  \bibinfo{author}{{Colpi}, M.}
\newblock \bibinfo{title}{{What if young z > 9 JWST galaxies hosted massive
  black holes?}}
\newblock \emph{\bibinfo{journal}{\mnras}} \textbf{\bibinfo{volume}{521}}~(1),
  \bibinfo{pages}{241--250} (\bibinfo{year}{2023}) .

\bibitem{zhu+2020}
\bibinfo{author}{{Zhu}, Q.} \emph{et~al.}
\newblock \bibinfo{title}{{The formation of the first quasars: the black hole
  seeds, accretion, and feedback models}}.
\newblock \emph{\bibinfo{journal}{\mnras}} \textbf{\bibinfo{volume}{514}}~(4),
  \bibinfo{pages}{5583--5606} (\bibinfo{year}{2022}) .

\bibitem{bhowmick+2022}
\bibinfo{author}{{Bhowmick}, A.~K.} \emph{et~al.}
\newblock \bibinfo{title}{{Probing the z {\ensuremath{\gtrsim}} 6 quasars in a
  universe with IllustrisTNG physics: impact of gas-based black hole seeding
  models}}.
\newblock \emph{\bibinfo{journal}{\mnras}} \textbf{\bibinfo{volume}{516}}~(1),
  \bibinfo{pages}{138--157} (\bibinfo{year}{2022}) .

\bibitem{Reines15}
\bibinfo{author}{{Reines}, A.~E.} \& \bibinfo{author}{{Volonteri}, M.}
\newblock \bibinfo{title}{{Relations between Central Black Hole Mass and Total
  Galaxy Stellar Mass in the Local Universe}}.
\newblock \emph{\bibinfo{journal}{\apj}} \textbf{\bibinfo{volume}{813}}~(2),
  \bibinfo{pages}{82} (\bibinfo{year}{2015}) .

\bibitem{izumi_subaru_2019}
\bibinfo{author}{Izumi, T.} \emph{et~al.}
\newblock \bibinfo{title}{Subaru high-z exploration of low-luminosity quasars
  ({SHELLQs}). {VIII}. a less biased view of the early co-evolution of black
  holes and host galaxies} \textbf{\bibinfo{volume}{71}}, \bibinfo{pages}{111}
  .

\bibitem{matsuoka+2017}
\bibinfo{author}{{Matsuoka}, K.} \emph{et~al.}
\newblock \bibinfo{title}{{Chemical enrichment and accretion of nitrogen-loud
  quasars}}.
\newblock \emph{\bibinfo{journal}{\aap}} \textbf{\bibinfo{volume}{608}},
  \bibinfo{pages}{A90} (\bibinfo{year}{2017}) .

\bibitem{Koptelova2022}
\bibinfo{author}{{Koptelova}, E.} \& \bibinfo{author}{{Hwang}, C.-Y.}
\newblock \bibinfo{title}{{Dense nitrogen-enriched circumnuclear region of the
  new high-redshift quasar ULAS J0816+2134 at z=7.46}}.
\newblock \emph{\bibinfo{journal}{arXiv e-prints}}
  \bibinfo{pages}{arXiv:2212.05862} (\bibinfo{year}{2022}) .

\bibitem{ubler+2023}
\bibinfo{author}{{{\"U}bler}, H.} \emph{et~al.}
\newblock \bibinfo{title}{{GA-NIFS: A massive black hole in a low-metallicity
  AGN at z {\ensuremath{\sim}} 5.55 revealed by JWST/NIRSpec IFS}}.
\newblock \emph{\bibinfo{journal}{\aap}} \textbf{\bibinfo{volume}{677}},
  \bibinfo{pages}{A145} (\bibinfo{year}{2023}) .

\bibitem{Jiang08}
\bibinfo{author}{{Jiang}, L.}, \bibinfo{author}{{Fan}, X.} \&
  \bibinfo{author}{{Vestergaard}, M.}
\newblock \bibinfo{title}{{A Sample of Quasars with Strong Nitrogen Emission
  Lines from the Sloan Digital Sky Survey}}.
\newblock \emph{\bibinfo{journal}{\apj}} \textbf{\bibinfo{volume}{679}}~(2),
  \bibinfo{pages}{962--966} (\bibinfo{year}{2008}) .

\bibitem{Osterbrock06}
\bibinfo{author}{{Osterbrock}, D.~E.} \& \bibinfo{author}{{Ferland}, G.~J.}
\newblock \emph{\bibinfo{title}{{Astrophysics of gaseous nebulae and active
  galactic nuclei}}}  (\bibinfo{year}{2006}).

\bibitem{Mandal21}
\bibinfo{author}{{Mandal}, A.~K.} \emph{et~al.}
\newblock \bibinfo{title}{{Estimation of the size and structure of the broad
  line region using Bayesian approach}}.
\newblock \emph{\bibinfo{journal}{\mnras}} \textbf{\bibinfo{volume}{502}}~(2),
  \bibinfo{pages}{2140--2157} (\bibinfo{year}{2021}) .

\bibitem{volpato+2023}
\bibinfo{author}{{Volpato}, G.} \emph{et~al.}
\newblock \bibinfo{title}{{A Study of Primordial Very Massive Star Evolution}}.
\newblock \emph{\bibinfo{journal}{\apj}} \textbf{\bibinfo{volume}{944}}~(1),
  \bibinfo{pages}{40} (\bibinfo{year}{2023}) .

\bibitem{Eisenstein23}
\bibinfo{author}{{Eisenstein}, D.~J.} \emph{et~al.}
\newblock \bibinfo{title}{{Overview of the JWST Advanced Deep Extragalactic
  Survey (JADES)}}.
\newblock \emph{\bibinfo{journal}{arXiv e-prints}}
  \bibinfo{pages}{arXiv:2306.02465} (\bibinfo{year}{2023}) .

\bibitem{Jakobsen22}
\bibinfo{author}{{Jakobsen}, P.} \emph{et~al.}
\newblock \bibinfo{title}{{The Near-Infrared Spectrograph (NIRSpec) on the
  James Webb Space Telescope. I. Overview of the instrument and its
  capabilities}}.
\newblock \emph{\bibinfo{journal}{\aap}} \textbf{\bibinfo{volume}{661}},
  \bibinfo{pages}{A80} (\bibinfo{year}{2022}) .

\bibitem{Ferruit22}
\bibinfo{author}{{Ferruit}, P.} \emph{et~al.}
\newblock \bibinfo{title}{{The Near-Infrared Spectrograph (NIRSpec) on the
  James Webb Space Telescope. II. Multi-object spectroscopy (MOS)}}.
\newblock \emph{\bibinfo{journal}{\aap}} \textbf{\bibinfo{volume}{661}},
  \bibinfo{pages}{A81} (\bibinfo{year}{2022}) .

\bibitem{Boker23}
\bibinfo{author}{{B{\"o}ker}, T.} \emph{et~al.}
\newblock \bibinfo{title}{{In-orbit Performance of the Near-infrared
  Spectrograph NIRSpec on the James Webb Space Telescope}}.
\newblock \emph{\bibinfo{journal}{\pasp}}
  \textbf{\bibinfo{volume}{135}}~(1045), \bibinfo{pages}{038001}
  (\bibinfo{year}{2023}) .

\bibitem{bushouse_2023}
\bibinfo{author}{{Bushouse}, H.} \emph{et~al.}
\newblock \bibinfo{title}{{JWST Calibration Pipeline}}  (\bibinfo{year}{2023})
  .

\bibitem{Planck18}
\bibinfo{author}{{Planck Collaboration}} \emph{et~al.}
\newblock \bibinfo{title}{{Planck 2018 results. VI. Cosmological parameters}}.
\newblock \emph{\bibinfo{journal}{\aap}} \textbf{\bibinfo{volume}{641}},
  \bibinfo{pages}{A6} (\bibinfo{year}{2020}) .

\bibitem{Maiolino23b}
\bibinfo{author}{{Maiolino}, R.} \emph{et~al.}
\newblock \bibinfo{title}{{JWST-JADES. Possible Population III signatures at
  z=10.6 in the halo of GN-z11}}.
\newblock \emph{\bibinfo{journal}{arXiv e-prints}}
  \bibinfo{pages}{arXiv:2306.00953} (\bibinfo{year}{2023}) .

\bibitem{Laha21}
\bibinfo{author}{{Laha}, S.} \emph{et~al.}
\newblock \bibinfo{title}{{Ionized outflows from active galactic nuclei as the
  essential elements of feedback}}.
\newblock \emph{\bibinfo{journal}{Nature Astronomy}}
  \textbf{\bibinfo{volume}{5}}, \bibinfo{pages}{13--24} (\bibinfo{year}{2021})
  .

\bibitem{rodriguez-hidalgo2009}
\bibinfo{author}{{Rodr{\'\i}guez Hidalgo}, P.}
\newblock \emph{\bibinfo{title}{{High velocity outflows in quasars}}}.
\newblock Ph.D. thesis, \bibinfo{school}{University of Florida}
  (\bibinfo{year}{2009}).

\bibitem{Hamann04}
\bibinfo{author}{{Hamann}, F.} \& \bibinfo{author}{{Sabra}, B.}
\newblock \bibinfo{editor}{{Richards}, G.~T.} \& \bibinfo{editor}{{Hall},
  P.~B.} (eds) \emph{\bibinfo{title}{{The Diverse Nature of Intrinsic Absorbers
  in AGNs}}}.
\newblock (eds \bibinfo{editor}{{Richards}, G.~T.} \& \bibinfo{editor}{{Hall},
  P.~B.}) \emph{\bibinfo{booktitle}{AGN Physics with the Sloan Digital Sky
  Survey}}, Vol. \bibinfo{volume}{311} of \emph{\bibinfo{series}{Astronomical
  Society of the Pacific Conference Series}}, \bibinfo{pages}{203}
  (\bibinfo{year}{2004}).

\bibitem{Chen17}
\bibinfo{author}{{Chen}, Z.-F.} \& \bibinfo{author}{{Pan}, D.-S.}
\newblock \bibinfo{title}{{Collective Properties of Quasar Narrow Associated
  Absorption Lines}}.
\newblock \emph{\bibinfo{journal}{\apj}} \textbf{\bibinfo{volume}{848}}~(2),
  \bibinfo{pages}{79} (\bibinfo{year}{2017}) .

\bibitem{Doyee22}
\bibinfo{author}{{Byun}, D.}, \bibinfo{author}{{Arav}, N.} \&
  \bibinfo{author}{{Hall}, P.~B.}
\newblock \bibinfo{title}{{The Farthest Quasar Mini-Broad Absorption Line
  Outflow from Its Central Source: Very Large Telescope/UVES Observation of
  SDSS J0242+0049}}.
\newblock \emph{\bibinfo{journal}{\apj}} \textbf{\bibinfo{volume}{927}}~(2),
  \bibinfo{pages}{176} (\bibinfo{year}{2022}) .

\bibitem{Pettini02}
\bibinfo{author}{{Pettini}, M.} \emph{et~al.}
\newblock \bibinfo{title}{{New Observations of the Interstellar Medium in the
  Lyman Break Galaxy MS 1512-cB58}}.
\newblock \emph{\bibinfo{journal}{\apj}} \textbf{\bibinfo{volume}{569}}~(2),
  \bibinfo{pages}{742--757} (\bibinfo{year}{2002}) .

\bibitem{bischetti+2022}
\bibinfo{author}{{Bischetti}, M.} \emph{et~al.}
\newblock \bibinfo{title}{{Suppression of black-hole growth by strong outflows
  at redshifts 5.8-6.6}}.
\newblock \emph{\bibinfo{journal}{\nat}} \textbf{\bibinfo{volume}{605}}~(7909),
  \bibinfo{pages}{244--247} (\bibinfo{year}{2022}) .

\bibitem{maiolino+2004}
\bibinfo{author}{{Maiolino}, R.} \emph{et~al.}
\newblock \bibinfo{title}{{Extreme gas properties in the most distant
  quasars}}.
\newblock \emph{\bibinfo{journal}{\aap}} \textbf{\bibinfo{volume}{420}},
  \bibinfo{pages}{889--897} (\bibinfo{year}{2004}) .

\bibitem{Diamond-Stanic12}
\bibinfo{author}{{Diamond-Stanic}, A.~M.} \emph{et~al.}
\newblock \bibinfo{title}{{High-velocity Outflows without AGN Feedback:
  Eddington-limited Star Formation in Compact Massive Galaxies}}.
\newblock \emph{\bibinfo{journal}{\apjl}} \textbf{\bibinfo{volume}{755}}~(2),
  \bibinfo{pages}{L26} (\bibinfo{year}{2012}) .

\bibitem{Diamond-Stanic21}
\bibinfo{author}{{Diamond-Stanic}, A.~M.} \emph{et~al.}
\newblock \bibinfo{title}{{Compact Starburst Galaxies with Fast Outflows:
  Central Escape Velocities and Stellar Mass Surface Densities from Multiband
  Hubble Space Telescope Imaging}}.
\newblock \emph{\bibinfo{journal}{\apj}} \textbf{\bibinfo{volume}{912}}~(1),
  \bibinfo{pages}{11} (\bibinfo{year}{2021}) .

\bibitem{Sell14}
\bibinfo{author}{{Sell}, P.~H.} \emph{et~al.}
\newblock \bibinfo{title}{{Massive compact galaxies with high-velocity
  outflows: morphological analysis and constraints on AGN activity}}.
\newblock \emph{\bibinfo{journal}{\mnras}} \textbf{\bibinfo{volume}{441}}~(4),
  \bibinfo{pages}{3417--3443} (\bibinfo{year}{2014}) .

\bibitem{Perrotta21}
\bibinfo{author}{{Perrotta}, S.} \emph{et~al.}
\newblock \bibinfo{title}{{Physical Properties of Massive Compact Starburst
  Galaxies with Extreme Outflows}}.
\newblock \emph{\bibinfo{journal}{\apj}} \textbf{\bibinfo{volume}{923}}~(2),
  \bibinfo{pages}{275} (\bibinfo{year}{2021}) .

\bibitem{arav+2018}
\bibinfo{author}{{Arav}, N.} \emph{et~al.}
\newblock \bibinfo{title}{{Evidence that 50\% of BALQSO Outflows Are Situated
  at Least 100 pc from the Central Source}}.
\newblock \emph{\bibinfo{journal}{\apj}} \textbf{\bibinfo{volume}{857}}~(1),
  \bibinfo{pages}{60} (\bibinfo{year}{2018}) .

\bibitem{xu+2019}
\bibinfo{author}{{Xu}, X.}, \bibinfo{author}{{Arav}, N.},
  \bibinfo{author}{{Miller}, T.} \& \bibinfo{author}{{Benn}, C.}
\newblock \bibinfo{title}{{VLT/X-Shooter Survey of BAL Quasars: Large Distance
  Scale and AGN Feedback}}.
\newblock \emph{\bibinfo{journal}{\apj}} \textbf{\bibinfo{volume}{876}}~(2),
  \bibinfo{pages}{105} (\bibinfo{year}{2019}) .

\bibitem{byun+2022}
\bibinfo{author}{{Byun}, D.}, \bibinfo{author}{{Arav}, N.} \&
  \bibinfo{author}{{Hall}, P.~B.}
\newblock \bibinfo{title}{{The Farthest Quasar Mini-Broad Absorption Line
  Outflow from Its Central Source: Very Large Telescope/UVES Observation of
  SDSS J0242+0049}}.
\newblock \emph{\bibinfo{journal}{\apj}} \textbf{\bibinfo{volume}{927}}~(2),
  \bibinfo{pages}{176} (\bibinfo{year}{2022}) .

\bibitem{walker+2022}
\bibinfo{author}{{Walker}, A.}, \bibinfo{author}{{Arav}, N.} \&
  \bibinfo{author}{{Byun}, D.}
\newblock \bibinfo{title}{{High mass flow rate in a BAL outflow of quasar SDSS
  J1130 + 0411}}.
\newblock \emph{\bibinfo{journal}{\mnras}} \textbf{\bibinfo{volume}{516}}~(3),
  \bibinfo{pages}{3778--3785} (\bibinfo{year}{2022}) .

\bibitem{Arav18}
\bibinfo{author}{{Arav}, N.} \emph{et~al.}
\newblock \bibinfo{title}{{Evidence that 50\% of BALQSO Outflows Are Situated
  at Least 100 pc from the Central Source}}.
\newblock \emph{\bibinfo{journal}{\apj}} \textbf{\bibinfo{volume}{857}}~(1),
  \bibinfo{pages}{60} (\bibinfo{year}{2018}) .

\bibitem{Chisholm20}
\bibinfo{author}{{Chisholm}, J.}, \bibinfo{author}{{Prochaska}, J.~X.},
  \bibinfo{author}{{Schaerer}, D.}, \bibinfo{author}{{Gazagnes}, S.} \&
  \bibinfo{author}{{Henry}, A.}
\newblock \bibinfo{title}{{Optically thin spatially resolved Mg II emission
  maps the escape of ionizing photons}}.
\newblock \emph{\bibinfo{journal}{\mnras}} \textbf{\bibinfo{volume}{498}}~(2),
  \bibinfo{pages}{2554--2574} (\bibinfo{year}{2020}) .

\bibitem{kuraszkiewicz+2004}
\bibinfo{author}{{Kuraszkiewicz}, J.~K.} \emph{et~al.}
\newblock \bibinfo{title}{{Emission Line Properties of Active Galactic Nuclei
  from a Post-COSTAR Hubble Space Telescope Faint Object Spectrograph Spectral
  Atlas}}.
\newblock \emph{\bibinfo{journal}{\apjs}} \textbf{\bibinfo{volume}{150}}~(1),
  \bibinfo{pages}{165--180} (\bibinfo{year}{2004}) .

\bibitem{nagao+2006b}
\bibinfo{author}{{Nagao}, T.}, \bibinfo{author}{{Maiolino}, R.} \&
  \bibinfo{author}{{Marconi}, A.}
\newblock \bibinfo{title}{{Gas metallicity in the narrow-line regions of
  high-redshift active galactic nuclei}}.
\newblock \emph{\bibinfo{journal}{\aap}} \textbf{\bibinfo{volume}{447}}~(3),
  \bibinfo{pages}{863--876} (\bibinfo{year}{2006}) .

\bibitem{cleri+2023}
\bibinfo{author}{{Cleri}, N.~J.} \emph{et~al.}
\newblock \bibinfo{title}{{Using [Ne V]/[Ne III] to Understand the Nature of
  Extreme-Ionization Galaxies}}.
\newblock \emph{\bibinfo{journal}{arXiv e-prints}}
  \bibinfo{pages}{arXiv:2301.07745} (\bibinfo{year}{2023}) .

\bibitem{vanzella+2010}
\bibinfo{author}{{Vanzella}, E.} \emph{et~al.}
\newblock \bibinfo{title}{{The unusual N IV] -emitter galaxy GDS
  J033218.92-275302.7: star formation or AGN-driven winds from a massive galaxy
  at z = 5.56}}.
\newblock \emph{\bibinfo{journal}{\aap}} \textbf{\bibinfo{volume}{513}},
  \bibinfo{pages}{A20} (\bibinfo{year}{2010}) .

\bibitem{Glikman07}
\bibinfo{author}{{Glikman}, E.}, \bibinfo{author}{{Djorgovski}, S.~G.},
  \bibinfo{author}{{Stern}, D.}, \bibinfo{author}{{Bogosavljevi{\'c}}, M.} \&
  \bibinfo{author}{{Mahabal}, A.}
\newblock \bibinfo{title}{{Discovery of Two Spectroscopically Peculiar,
  Low-Luminosity Quasars at z\raisebox{-0.5ex}\textasciitilde4}}.
\newblock \emph{\bibinfo{journal}{\apjl}} \textbf{\bibinfo{volume}{663}}~(2),
  \bibinfo{pages}{L73--L76} (\bibinfo{year}{2007}) .

\bibitem{Gutkin16}
\bibinfo{author}{{Gutkin}, J.}, \bibinfo{author}{{Charlot}, S.} \&
  \bibinfo{author}{{Bruzual}, G.}
\newblock \bibinfo{title}{{Modelling the nebular emission from primeval to
  present-day star-forming galaxies}}.
\newblock \emph{\bibinfo{journal}{\mnras}} \textbf{\bibinfo{volume}{462}}~(2),
  \bibinfo{pages}{1757--1774} (\bibinfo{year}{2016}) .

\bibitem{Nakajima22}
\bibinfo{author}{{Nakajima}, K.} \& \bibinfo{author}{{Maiolino}, R.}
\newblock \bibinfo{title}{{Diagnostics for PopIII galaxies and direct collapse
  black holes in the early universe}}.
\newblock \emph{\bibinfo{journal}{\mnras}} \textbf{\bibinfo{volume}{513}}~(4),
  \bibinfo{pages}{5134--5147} (\bibinfo{year}{2022}) .

\bibitem{Ferland17}
\bibinfo{author}{{Ferland}, G.~J.} \emph{et~al.}
\newblock \bibinfo{title}{{The 2017 Release Cloudy}}.
\newblock \emph{\bibinfo{journal}{\rmxaa}} \textbf{\bibinfo{volume}{53}},
  \bibinfo{pages}{385--438} (\bibinfo{year}{2017}) .

\bibitem{Eldridge17}
\bibinfo{author}{{Eldridge}, J.~J.} \emph{et~al.}
\newblock \bibinfo{title}{{Binary Population and Spectral Synthesis Version
  2.1: Construction, Observational Verification, and New Results}}.
\newblock \emph{\bibinfo{journal}{\pasa}} \textbf{\bibinfo{volume}{34}},
  \bibinfo{pages}{e058} (\bibinfo{year}{2017}) .

\bibitem{Salpeter55}
\bibinfo{author}{{Salpeter}, E.~E.}
\newblock \bibinfo{title}{{The Luminosity Function and Stellar Evolution.}}
\newblock \emph{\bibinfo{journal}{\apj}} \textbf{\bibinfo{volume}{121}},
  \bibinfo{pages}{161} (\bibinfo{year}{1955}) .

\bibitem{shakura+sunyaev1973}
\bibinfo{author}{{Shakura}, N.~I.} \& \bibinfo{author}{{Sunyaev}, R.~A.}
\newblock \bibinfo{title}{{Black holes in binary systems. Observational
  appearance.}}
\newblock \emph{\bibinfo{journal}{\aap}} \textbf{\bibinfo{volume}{24}},
  \bibinfo{pages}{337--355} (\bibinfo{year}{1973}) .

\bibitem{capellupo+2015}
\bibinfo{author}{{Capellupo}, D.~M.}, \bibinfo{author}{{Netzer}, H.},
  \bibinfo{author}{{Lira}, P.}, \bibinfo{author}{{Trakhtenbrot}, B.} \&
  \bibinfo{author}{{Mej{\'\i}a-Restrepo}, J.}
\newblock \bibinfo{title}{{Active galactic nuclei at z {\ensuremath{\sim}} 1.5
  - I. Spectral energy distribution and accretion discs}}.
\newblock \emph{\bibinfo{journal}{\mnras}} \textbf{\bibinfo{volume}{446}}~(4),
  \bibinfo{pages}{3427--3446} (\bibinfo{year}{2015}) .

\bibitem{leighly+moore2004}
\bibinfo{author}{{Leighly}, K.~M.} \& \bibinfo{author}{{Moore}, J.~R.}
\newblock \bibinfo{title}{{Hubble Space Telescope STIS Ultraviolet Spectral
  Evidence of Outflow in Extreme Narrow-Line Seyfert 1 Galaxies. I. Data and
  Analysis}}.
\newblock \emph{\bibinfo{journal}{\apj}} \textbf{\bibinfo{volume}{611}}~(1),
  \bibinfo{pages}{107--124} (\bibinfo{year}{2004}) .

\bibitem{Maiolino01a}
\bibinfo{author}{{Maiolino}, R.}, \bibinfo{author}{{Salvati}, M.},
  \bibinfo{author}{{Marconi}, A.} \& \bibinfo{author}{{Antonucci}, R.~R.~J.}
\newblock \bibinfo{title}{{The Ly-edge paradox and the need for obscured
  QSOs}}.
\newblock \emph{\bibinfo{journal}{\aap}} \textbf{\bibinfo{volume}{375}},
  \bibinfo{pages}{25--29} (\bibinfo{year}{2001}) .

\bibitem{Maiolino03}
\bibinfo{author}{{Maiolino}, R.}, \bibinfo{author}{{Juarez}, Y.},
  \bibinfo{author}{{Mujica}, R.}, \bibinfo{author}{{Nagar}, N.~M.} \&
  \bibinfo{author}{{Oliva}, E.}
\newblock \bibinfo{title}{{Early Star Formation Traced by the Highest Redshift
  Quasars}}.
\newblock \emph{\bibinfo{journal}{\apjl}} \textbf{\bibinfo{volume}{596}}~(2),
  \bibinfo{pages}{L155--L158} (\bibinfo{year}{2003}) .

\bibitem{DeRosa11}
\bibinfo{author}{{De Rosa}, G.} \emph{et~al.}
\newblock \bibinfo{title}{{Evidence for Non-evolving Fe II/Mg II Ratios in
  Rapidly Accreting z \raisebox{-0.5ex}\textasciitilde 6 QSOs}}.
\newblock \emph{\bibinfo{journal}{\apj}} \textbf{\bibinfo{volume}{739}}~(2),
  \bibinfo{pages}{56} (\bibinfo{year}{2011}) .

\bibitem{Mazzucchelli17}
\bibinfo{author}{{Mazzucchelli}, C.} \emph{et~al.}
\newblock \bibinfo{title}{{Physical Properties of 15 Quasars at z
  {\ensuremath{\gtrsim}} 6.5}}.
\newblock \emph{\bibinfo{journal}{\apj}} \textbf{\bibinfo{volume}{849}}~(2),
  \bibinfo{pages}{91} (\bibinfo{year}{2017}) .

\bibitem{Shin19}
\bibinfo{author}{{Shin}, J.}, \bibinfo{author}{{Nagao}, T.},
  \bibinfo{author}{{Woo}, J.-H.} \& \bibinfo{author}{{Le}, H. A.~N.}
\newblock \bibinfo{title}{{The Fe II/Mg II Flux Ratio of Low-luminosity Quasars
  at z {\ensuremath{\sim}} 3}}.
\newblock \emph{\bibinfo{journal}{\apj}} \textbf{\bibinfo{volume}{874}}~(1),
  \bibinfo{pages}{22} (\bibinfo{year}{2019}) .

\bibitem{Sameshima20}
\bibinfo{author}{{Sameshima}, H.} \emph{et~al.}
\newblock \bibinfo{title}{{Mg II and Fe II Fluxes of Luminous Quasars at z
  {\ensuremath{\sim}} 2.7 and the Evaluation of the Baldwin Effect in the
  Flux-to-abundance Conversion Method for Quasars}}.
\newblock \emph{\bibinfo{journal}{\apj}} \textbf{\bibinfo{volume}{904}}~(2),
  \bibinfo{pages}{162} (\bibinfo{year}{2020}) .

\bibitem{Maiolino19}
\bibinfo{author}{{Maiolino}, R.} \& \bibinfo{author}{{Mannucci}, F.}
\newblock \bibinfo{title}{{De re metallica: the cosmic chemical evolution of
  galaxies}}.
\newblock \emph{\bibinfo{journal}{\aapr}} \textbf{\bibinfo{volume}{27}}~(1),
  \bibinfo{pages}{3} (\bibinfo{year}{2019}) .

\bibitem{oesch+2014}
\bibinfo{author}{{Oesch}, P.~A.} \emph{et~al.}
\newblock \bibinfo{title}{{The Most Luminous z \raisebox{-0.5ex}\textasciitilde
  9-10 Galaxy Candidates Yet Found: The Luminosity Function, Cosmic
  Star-formation Rate, and the First Mass Density Estimate at 500 Myr}}.
\newblock \emph{\bibinfo{journal}{\apj}} \textbf{\bibinfo{volume}{786}}~(2),
  \bibinfo{pages}{108} (\bibinfo{year}{2014}) .

\bibitem{oesch_remarkably_2016}
\bibinfo{author}{Oesch, P.~A.} \emph{et~al.}
\newblock \bibinfo{title}{A remarkably luminous galaxy at z=11.1 measured with
  hubble space telescope grism spectroscopy}.
\newblock \emph{\bibinfo{journal}{\apj}} \textbf{\bibinfo{volume}{819}},
  \bibinfo{pages}{129} (\bibinfo{year}{2016}) .

\bibitem{Ai13}
\bibinfo{author}{{Ai}, Y.~L.} \emph{et~al.}
\newblock \bibinfo{title}{{A Comparative Study of Optical/Ultraviolet
  Variability of Narrow-line Seyfert 1 and Broad-line Seyfert 1 Active Galactic
  Nuclei}}.
\newblock \emph{\bibinfo{journal}{\aj}} \textbf{\bibinfo{volume}{145}}~(4),
  \bibinfo{pages}{90} (\bibinfo{year}{2013}) .

\bibitem{Xue16}
\bibinfo{author}{{Xue}, Y.~Q.} \emph{et~al.}
\newblock \bibinfo{title}{{The 2 Ms Chandra Deep Field-North Survey and the 250
  ks Extended Chandra Deep Field-South Survey: Improved Point-source
  Catalogs}}.
\newblock \emph{\bibinfo{journal}{\apjs}} \textbf{\bibinfo{volume}{224}}~(2),
  \bibinfo{pages}{15} (\bibinfo{year}{2016}) .

\bibitem{Vasudevan07}
\bibinfo{author}{{Vasudevan}, R.~V.} \& \bibinfo{author}{{Fabian}, A.~C.}
\newblock \bibinfo{title}{{Piecing together the X-ray background: bolometric
  corrections for active galactic nuclei}}.
\newblock \emph{\bibinfo{journal}{\mnras}} \textbf{\bibinfo{volume}{381}}~(3),
  \bibinfo{pages}{1235--1251} (\bibinfo{year}{2007}) .

\bibitem{Buisson18}
\bibinfo{author}{{Buisson}, D.~J.~K.}, \bibinfo{author}{{Fabian}, A.~C.} \&
  \bibinfo{author}{{Lohfink}, A.~M.}
\newblock \bibinfo{title}{{Coronal temperatures of the AGN ESO 103-035 and IGR
  2124.7+5058 from NuSTAR observations}}.
\newblock \emph{\bibinfo{journal}{\mnras}} \textbf{\bibinfo{volume}{481}}~(4),
  \bibinfo{pages}{4419--4426} (\bibinfo{year}{2018}) .

\bibitem{Netzer19}
\bibinfo{author}{{Netzer}, H.}
\newblock \bibinfo{title}{{Bolometric correction factors for active galactic
  nuclei}}.
\newblock \emph{\bibinfo{journal}{\mnras}} \textbf{\bibinfo{volume}{488}}~(4),
  \bibinfo{pages}{5185--5191} (\bibinfo{year}{2019}) .

\bibitem{Willott10}
\bibinfo{author}{{Willott}, C.~J.} \emph{et~al.}
\newblock \bibinfo{title}{{Eddington-limited Accretion and the Black Hole Mass
  Function at Redshift 6}}.
\newblock \emph{\bibinfo{journal}{\aj}} \textbf{\bibinfo{volume}{140}}~(2),
  \bibinfo{pages}{546--560} (\bibinfo{year}{2010}) .

\bibitem{Trakhtenbrot11}
\bibinfo{author}{{Trakhtenbrot}, B.}, \bibinfo{author}{{Netzer}, H.},
  \bibinfo{author}{{Lira}, P.} \& \bibinfo{author}{{Shemmer}, O.}
\newblock \bibinfo{title}{{Black Hole Mass and Growth Rate at z
  \raisebox{-0.5ex}\textasciitilde= 4.8: A Short Episode of Fast Growth
  Followed by Short Duty Cycle Activity}}.
\newblock \emph{\bibinfo{journal}{\apj}} \textbf{\bibinfo{volume}{730}}~(1),
  \bibinfo{pages}{7} (\bibinfo{year}{2011}) .

\bibitem{Shen11}
\bibinfo{author}{{Shen}, Y.} \emph{et~al.}
\newblock \bibinfo{title}{{A Catalog of Quasar Properties from Sloan Digital
  Sky Survey Data Release 7}}.
\newblock \emph{\bibinfo{journal}{\apjs}} \textbf{\bibinfo{volume}{194}}~(2),
  \bibinfo{pages}{45} (\bibinfo{year}{2011}) .

\bibitem{Shen19}
\bibinfo{author}{{Shen}, Y.} \emph{et~al.}
\newblock \bibinfo{title}{{Gemini GNIRS Near-infrared Spectroscopy of 50
  Quasars at z {\ensuremath{\gtrsim}} 5.7}}.
\newblock \emph{\bibinfo{journal}{\apj}} \textbf{\bibinfo{volume}{873}}~(1),
  \bibinfo{pages}{35} (\bibinfo{year}{2019}) .

\bibitem{Pensabene20}
\bibinfo{author}{{Pensabene}, A.} \emph{et~al.}
\newblock \bibinfo{title}{{The ALMA view of the high-redshift relation between
  supermassive black holes and their host galaxies}}.
\newblock \emph{\bibinfo{journal}{\aap}} \textbf{\bibinfo{volume}{637}},
  \bibinfo{pages}{A84} (\bibinfo{year}{2020}) .

\bibitem{izumi_subaru_2021}
\bibinfo{author}{Izumi, T.} \emph{et~al.}
\newblock \bibinfo{title}{Subaru high-z exploration of low-luminosity quasars
  ({SHELLQs}). {XIII}. large-scale feedback and star formation in a
  low-luminosity quasar at z = 7.07 on the local black hole to host mass
  relation}.
\newblock \emph{\bibinfo{journal}{\apj}} \textbf{\bibinfo{volume}{914}},
  \bibinfo{pages}{36} (\bibinfo{year}{2021}) .

\bibitem{Harikane23}
\bibinfo{author}{{Harikane}, Y.} \emph{et~al.}
\newblock \bibinfo{title}{{JWST/NIRSpec First Census of Broad-Line AGNs at
  z=4-7: Detection of 10 Faint AGNs with
  M\_BH\raisebox{-0.5ex}\textasciitilde10\^6-10\^7 M\_sun and Their Host Galaxy
  Properties}}.
\newblock \emph{\bibinfo{journal}{arXiv e-prints}}
  \bibinfo{pages}{arXiv:2303.11946} (\bibinfo{year}{2023}) .

\bibitem{Mezcua23}
\bibinfo{author}{{Mezcua}, M.} \emph{et~al.}
\newblock \bibinfo{title}{{Overmassive Black Holes in Dwarf Galaxies Out to z
  0.9 in the VIPERS Survey}}.
\newblock \emph{\bibinfo{journal}{\apjl}} \textbf{\bibinfo{volume}{943}}~(1),
  \bibinfo{pages}{L5} (\bibinfo{year}{2023}) .

\bibitem{Vestergaard09}
\bibinfo{author}{{Vestergaard}, M.} \& \bibinfo{author}{{Osmer}, P.~S.}
\newblock \bibinfo{title}{{Mass Functions of the Active Black Holes in Distant
  Quasars from the Large Bright Quasar Survey, the Bright Quasar Survey, and
  the Color-selected Sample of the SDSS Fall Equatorial Stripe}}.
\newblock \emph{\bibinfo{journal}{\apj}} \textbf{\bibinfo{volume}{699}}~(1),
  \bibinfo{pages}{800--816} (\bibinfo{year}{2009}) .

\bibitem{Kormendy13}
\bibinfo{author}{{Kormendy}, J.} \& \bibinfo{author}{{Ho}, L.~C.}
\newblock \bibinfo{title}{{Coevolution (Or Not) of Supermassive Black Holes and
  Host Galaxies}}.
\newblock \emph{\bibinfo{journal}{\araa}} \textbf{\bibinfo{volume}{51}}~(1),
  \bibinfo{pages}{511--653} (\bibinfo{year}{2013}) .

\bibitem{Maiolino23c}
\bibinfo{author}{{Maiolino}, R.} \emph{et~al.}
\newblock \bibinfo{title}{{JADES. The diverse population of infant Black Holes
  at 4<z<11: merging, tiny, poor, but mighty}}.
\newblock \emph{\bibinfo{journal}{arXiv e-prints}}
  \bibinfo{pages}{arXiv:2308.01230} (\bibinfo{year}{2023}) .

\bibitem{Marconi2008}
\bibinfo{author}{{Marconi}, A.} \emph{et~al.}
\newblock \bibinfo{title}{{The Effect of Radiation Pressure on Virial Black
  Hole Mass Estimates and the Case of Narrow-Line Seyfert 1 Galaxies}}.
\newblock \emph{\bibinfo{journal}{\apj}} \textbf{\bibinfo{volume}{678}}~(2),
  \bibinfo{pages}{693--700} (\bibinfo{year}{2008}) .

\bibitem{Khatu23}
\bibinfo{author}{{Khatu}, V.~C.} \emph{et~al.}
\newblock \bibinfo{title}{{Supermassive Black Holes with High Accretion Rates
  in Active Galactic Nuclei. XIII. Ultraviolet Time Lag of H$\beta$ Emission in
  Mrk 142}}.
\newblock \emph{\bibinfo{journal}{arXiv e-prints}}
  \bibinfo{pages}{arXiv:2309.13418} (\bibinfo{year}{2023}) .

\bibitem{Kocevski23}
\bibinfo{author}{{Kocevski}, D.~D.} \emph{et~al.}
\newblock \bibinfo{title}{{Hidden Little Monsters: Spectroscopic Identification
  of Low-Mass, Broad-Line AGN at $z>5$ with CEERS}}.
\newblock \emph{\bibinfo{journal}{arXiv e-prints}}
  \bibinfo{pages}{arXiv:2302.00012} (\bibinfo{year}{2023}) .

\bibitem{Ding22}
\bibinfo{author}{{Ding}, X.} \emph{et~al.}
\newblock \bibinfo{title}{{First detections of stellar light from quasar host
  galaxies at z > 6}}.
\newblock \emph{\bibinfo{journal}{arXiv e-prints}}
  \bibinfo{pages}{arXiv:2211.14329} (\bibinfo{year}{2022}) .

\bibitem{Schotlz23}
\bibinfo{author}{{Scholtz}, J.} \emph{et~al.}
\newblock \bibinfo{title}{{GN-z11: The environment of an AGN at $z=$10.603}}.
\newblock \emph{\bibinfo{journal}{arXiv e-prints}}
  \bibinfo{pages}{arXiv:2306.09142} (\bibinfo{year}{2023}) .

\bibitem{Zhang_Trinity_23_I}
\bibinfo{author}{{Zhang}, H.} \emph{et~al.}
\newblock \bibinfo{title}{{TRINITY I: self-consistently modelling the dark
  matter halo-galaxy-supermassive black hole connection from z = 0-10}}.
\newblock \emph{\bibinfo{journal}{\mnras}} \textbf{\bibinfo{volume}{518}}~(2),
  \bibinfo{pages}{2123--2163} (\bibinfo{year}{2023}) .

\bibitem{smidt+2018}
\bibinfo{author}{{Smidt}, J.}, \bibinfo{author}{{Whalen}, D.~J.},
  \bibinfo{author}{{Johnson}, J.~L.}, \bibinfo{author}{{Surace}, M.} \&
  \bibinfo{author}{{Li}, H.}
\newblock \bibinfo{title}{{Radiation Hydrodynamical Simulations of the First
  Quasars}}.
\newblock \emph{\bibinfo{journal}{\apj}} \textbf{\bibinfo{volume}{865}}~(2),
  \bibinfo{pages}{126} (\bibinfo{year}{2018}) .

\bibitem{Trinca22}
\bibinfo{author}{{Trinca}, A.} \emph{et~al.}
\newblock \bibinfo{title}{{The low-end of the black hole mass function at
  cosmic dawn}}.
\newblock \emph{\bibinfo{journal}{\mnras}} \textbf{\bibinfo{volume}{511}}~(1),
  \bibinfo{pages}{616--640} (\bibinfo{year}{2022}) .

\bibitem{trinca+2023}
\bibinfo{author}{{Trinca}, A.} \emph{et~al.}
\newblock \bibinfo{title}{{Seeking the growth of the first black hole seeds
  with JWST}}.
\newblock \emph{\bibinfo{journal}{\mnras}} \textbf{\bibinfo{volume}{519}}~(3),
  \bibinfo{pages}{4753--4764} (\bibinfo{year}{2023}) .

\bibitem{Liu20}
\bibinfo{author}{{Liu}, B.} \& \bibinfo{author}{{Bromm}, V.}
\newblock \bibinfo{title}{{When did Population III star formation end?}}
\newblock \emph{\bibinfo{journal}{\mnras}} \textbf{\bibinfo{volume}{497}}~(3),
  \bibinfo{pages}{2839--2854} (\bibinfo{year}{2020}) .

\bibitem{Venditti23}
\bibinfo{author}{{Venditti}, A.} \emph{et~al.}
\newblock \bibinfo{title}{{A needle in a haystack? Catching Population III
  stars in the epoch of reionization: I. Population III star-forming
  environments}}.
\newblock \emph{\bibinfo{journal}{\mnras}} \textbf{\bibinfo{volume}{522}}~(3),
  \bibinfo{pages}{3809--3830} (\bibinfo{year}{2023}) .

\end{thebibliography}

\end{document}